\documentclass[runningheads]{llncs}
\usepackage[utf8]{inputenc}
\usepackage{nicefrac}
\usepackage{microtype}
\usepackage{amsmath}
\usepackage{amssymb}
\usepackage{xcolor}
\usepackage{listings}
\usepackage{subfigure}
\usepackage{tikz}
\usepackage{mdframed}
\usepackage{tikz}
\usepackage{wrapfig}
\usepackage{pgfplots}
\usepackage{graphicx}
\usepackage{tikz-cd}
\usepackage{paralist}
\usepackage{booktabs}
\usepackage{colonequals}
\usepackage{url}
\usepackage[firstpage]{draftwatermark}

\definecolor{shadecolor}{gray}{0.9}

\SetWatermarkText{\hspace*{5in}\raisebox{5.in}{\includegraphics[scale=0.8]{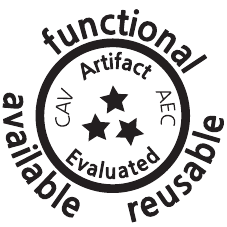}}}

\SetWatermarkAngle{0}

\usepackage[breaklinks=true]{hyperref}
\usepackage[inline]{enumitem}
\graphicspath{ {./figures/} }

\usetikzlibrary{shapes}

\usepackage{pifont}%

\makeatletter
\def\orcidID#1{\smash{\href{http://orcid.org/#1}{\protect\raisebox{-1.25pt}{\protect\includegraphics{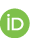}}}}}
\makeatother

\newcommand{\mysubsubsection}[1]{\medskip\noindent\textbf{#1}~~}
\newcommand{\myparagraph}[1]{\smallskip\noindent\emph{#1}}

\usetikzlibrary{patterns,arrows,backgrounds,calc,shapes,shadows,decorations.pathmorphing,decorations.pathreplacing,automata,shapes.multipart,positioning,shapes.geometric,fit,circuits,trees,shapes.gates.logic.US,fit,decorations.markings}

\tikzset{sstate/.style={circle, draw=black, inner sep=1pt}}
\pgfplotstableread[col sep = comma]{data/motiv-ex.csv}\motiv
\pgfplotstableread[col sep = comma]{data/herman-13.csv}\hermanthirteen 
\pgfplotstableread[col sep = comma]{data/herman-17.csv}\hermanseventeen
\pgfplotstableread[col sep = comma]{data/herman-19.csv}\hermannineteen
\pgfplotstableread[col sep = comma]{data/guy-mc.csv}\guymc
\pgfplotstableread[col sep = comma]{data/weather.csv}\weather
\pgfplotstableread[col sep = comma]{data/weather-2-state.csv}\weathertwo
\pgfplotstableread[col sep = comma]{data/queues.csv}\queues

\tikzstyle{nnf}=[
  >=stealth,font=\small,auto,scale=0.7,every node/.style={scale=0.7}
]
\tikzstyle{extnode}=[
  draw,circle,inner sep=2pt,fill=white
]

\tikzstyle{leafnode}=[
  draw,fill=gray!20,inner sep=3.5pt
]
\tikzstyle{constnode}=[
  draw,fill=white,inner sep=3.5pt
]
\tikzstyle{label}=[
  fill=white,inner sep=2.5pt
]

\tikzstyle{acarrow}=[
    decoration={markings,mark=at position 1 with {\arrow[scale=0.6]{>}}},
    postaction={decorate},
    shorten >=0.4pt,
    >=latex,
    line width=0.1
]

\tikzstyle{bnarrow}=[
    decoration={markings,mark=at position 1 with {\arrow[scale=1.5]{>}}},
    postaction={decorate},
    shorten >=0.7pt,
    >=latex,
    line width=0.3
]
\tikzstyle{bayesnet}=[
  >=latex, thick, auto
]
\tikzstyle{bnnode}=[
  draw,ellipse,minimum size=7mm,inner sep=1pt,font=\small
]
\tikzstyle{cpt}=[
  font=\footnotesize
]

\tikzstyle{graph}=[
  >=stealth,font=\small,auto,scale=1,every node/.style={scale=1}
]
\tikzstyle{node}=[
  draw,circle,inner sep=3pt,fill=white
]

\tikzstyle{bdd}=[
  >=latex, thick, >=stealth, font=\small,auto,scale=0.9,every node/.style={scale=0.9}
]
\tikzstyle{bddnode}=[
  draw,circle,inner sep=0pt,fill=white,minimum size=5.5mm
]

\tikzstyle{highedge}=[
    line width=0.9
]
\tikzstyle{lowedge}=[
    line width=0.9,dotted
]
\tikzstyle{bddterminal}=[
  draw,fill=gray!20,inner sep=2.5pt, font=\small
]

\definecolor{prismgreen}{HTML}{009900}
\definecolor{prismred}{HTML}{cc0000}
\definecolor{prismblue}{HTML}{0000cc}

\lstdefinelanguage{Prism}{
        basicstyle=\color{prismred}\scriptsize\ttfamily,
        literate=*	{0}{{\textcolor{prismblue}{0}}}{1}
			{1}{{\textcolor{prismblue}{1}}}{1}
			{2}{{\textcolor{prismblue}{2}}}{1}
			{3}{{\textcolor{prismblue}{3}}}{1}
			{4}{{\textcolor{prismblue}{4}}}{1}
			{5}{{\textcolor{prismblue}{5}}}{1}
			{6}{{\textcolor{prismblue}{6}}}{1}
			{7}{{\textcolor{prismblue}{7}}}{1}
			{8}{{\textcolor{prismblue}{8}}}{1}
			{9}{{\textcolor{prismblue}{9}}}{1}
			{.0}{{\textcolor{prismblue}{.0}}}{2}
			{.1}{{\textcolor{prismblue}{.1}}}{2}
			{.2}{{\textcolor{prismblue}{.2}}}{2}
			{.3}{{\textcolor{prismblue}{.3}}}{2}
			{.4}{{\textcolor{prismblue}{.4}}}{2}
			{.5}{{\textcolor{prismblue}{.5}}}{2}
			{.6}{{\textcolor{prismblue}{.6}}}{2}
			{.7}{{\textcolor{prismblue}{.7}}}{2}
			{.8}{{\textcolor{prismblue}{.8}}}{2}
			{.9}{{\textcolor{prismblue}{.9}}}{2}
			{[}{{\textcolor{black}{[}}}{1}
			{]}{{\textcolor{black}{]}}}{1}
			{+}{{\textcolor{black}{+}}}{1}
			{-}{{\textcolor{black}{-}}}{1}
			{=}{{\textcolor{black}{=}}}{1}
			{<}{{\textcolor{black}{<}}}{1}
			{>}{{\textcolor{black}{>}}}{1}
			{\&}{{\textcolor{black}{\&}}}{1}
			{|}{{\textcolor{black}{|}}}{1}
			{:}{{\textcolor{black}{:}}}{1}
			{;}{{\textcolor{black}{;}}}{1}
			{(}{{\textcolor{black}{(}}}{1}
			{)}{{\textcolor{black}{)}}}{1}
			{..}{{\textcolor{black}{..}}}{2},
        keywords= {bool,ceil,const,ctmc,double,dtmc,endinit,endmodule,endrewards, endsystem,F,false,floor,formula,G,global,I,init,int,label,max,mdp,min,module,nondeterministic,P,Pmin,Pmax,prob,probabilistic,rate,rewards,Rmin,Rmax,S,stochastic,system,true,U, option, either, assignment, relation, operation, hole, variable},
        keywordstyle={\bfseries\color{black}},
        numberstyle=\footnotesize\color{black},
        comment=[l] {//}, morecomment=[s]{/*}{*/},
        commentstyle= \color{prismgreen},
        tabsize=4,
        captionpos=b,
        escapechar=^,
        moredelim=[is][\color{orange}]{@}{@},
}%

\lstdefinelanguage{dice}{
  language=Caml,
  belowskip=0mm,
  showstringspaces=false,
  columns=flexible,
  basicstyle={\scriptsize\ttfamily},
  numberstyle=\color{gray},
  keywordstyle=\color{blue},
  commentstyle=\color{dkgreen},
  breaklines=true,
  tabsize=3,
}

\newcommand{\prism}{\textsc{Prism}}

\newcommand{\rubicon}{\textsc{Rubicon}}
\newcommand{\storm}{\textsc{Storm}}

\newcommand{\true}{\texttt{true}}

\newcommand{\dice}{\texttt{Dice}}
\newcommand{\readcmd}{\texttt{read}}
\newcommand{\writecmd}{\texttt{write}}
\newcommand{\BB}{\mathbb{B}}
\newcommand{\RR}{\mathbb{R}}
\newcommand{\QQ}{\mathbb{Q}}
\newcommand{\RRnn}{\mathbb{R}_{\geq 0}}

\newcommand{\prog}{\mathcal{P}}

\newcommand{\valmap}{\mathsf{val}}
\newcommand{\varmap}{\mathsf{var}} 
\newcommand{\weight}{\mathsf{weight}}
\newcommand{\program}[1][]{\mathcal{P}}
\newcommand{\semantics}[1]{[\![\, #1 \, ]\!]}
\newcommand{\globally}{\square}
\newcommand{\eventuallyb}[1]{\lozenge^{\leq #1}}
\newcommand{\eventually}{\lozenge}
\newcommand{\mc}{\mathcal{M}}
\newcommand{\Distr}{\ensuremath{Distr}}
\newcommand{\last}[1]{\ensuremath{#1_\downarrow}}
\newcommand{\ct}[2]{\ensuremath{\mathsf{CT}(#1,#2)}}
\newcommand{\Succ}{\ensuremath{\mathsf{Succ}}}
\newcommand{\bisim}{\sim}

\newcommand{\bdd}[2]{\mathsf{BDD}(#1, #2)}
\newcommand{\bddw}[3]{\mathsf{BDD}_\mathsf{MC}(#1, #2, #3)}
\newcommand{\coinsweight}{W}
\newcommand{\coins}{C}
\newcommand{\coin}{c}
\newcommand{\coinsorder}{{<}_\coins}
\newcommand{\WMC}{\mathsf{WMC}}
\newcommand{\seteventuallyb}[2]{\semantics{\eventuallyb{#1} #2}}
\newcommand{\seteventuallybfrom}[3]{\semantics{#1{\rightarrow}\eventuallyb{#2} #3}}

\newcommand{\hPaths}{\text{Paths}_h}
\newcommand{\bool}{\{0, 1\}}

\newcommand{\secref}[1]{Sec.~{\ref{sec:#1}}}

\newcommand{\repthanks}[1]{\textsuperscript{\ref{#1}}}
\makeatletter
\patchcmd{\maketitle}
  {\def\thanks}
  {\let\repthanks\repthanksunskip\def\thanks}
  {}{}
\patchcmd{\@maketitle}
  {\def\thanks}
  {\let\repthanks\@gobble\def\thanks}
  {}{}
\newcommand\repthanksunskip[1]{\unskip{}}
\makeatother

\title{Model Checking Finite-Horizon Markov Chains with Probabilistic Inference}
\author{Steven Holtzen\thanks{Contributed equally\protect\label{X}}\inst{1}\orcidID{0000-0002-8190-5412}\and Sebastian Junges\repthanks{X} \inst{2}\orcidID{0000-0003-0978-8466}\and Marcell Vazquez-Chanlatte\inst{2}\orcidID{0000-0002-1248-0000} \and \\Todd Millstein\inst{1}\orcidID{0000-0002-2031-1514} \and Sanjit A.\ Seshia\inst{2}\orcidID{0000-0001-6190-8707} \and Guy Van den Broeck\inst{1}\orcidID{0000-0003-3434-2503} 
}
\authorrunning{Holtzen \emph{et al.}}
\institute{
University of California, Los Angeles, CA, USA\thanks{This work is partially supported by NSF grants \#IIS-1943641, \#IIS-1956441, \#CCF-1837129, DARPA grant \#N66001-17-2-4032, a Sloan Fellowship, a UCLA Dissertation Year Fellowship, and gifts by Intel and Facebook Research.} \and 
University of California, Berkeley, CA, USA\thanks{This work is partially supported by NSF grants 1545126 (VeHICaL), 1646208 and 1837132, by the DARPA contracts FA8750-18-C-0101 (AA) and FA8750-20-C-0156 (SDCPS), by Berkeley Deep Drive, and by Toyota under the iCyPhy center.}
}

\begin{document}

\maketitle

\begin{abstract}
We revisit the symbolic verification of Markov chains with respect to finite horizon reachability properties. The prevalent approach iteratively computes step-bounded state reachability probabilities. 
By contrast, recent advances in probabilistic inference suggest symbolically representing all horizon-length paths through the Markov chain. We ask whether this perspective advances the state-of-the-art in probabilistic model checking.      
First, we formally describe both approaches in order to highlight their key differences.   
Then, using these insights we develop \rubicon{}, a tool that transpiles \prism{} models to the probabilistic inference tool \dice{}. 
Finally, we demonstrate better scalability compared to probabilistic model checkers on selected benchmarks. 
All together, our results suggest that probabilistic inference is a valuable addition to the probabilistic model checking portfolio, with \rubicon{} as a first step towards integrating both perspectives.

\end{abstract}

\section{Introduction}

Systems with probabilistic uncertainty are ubiquitous, e.g.,
probabilistic programs, distributed systems, fault trees,
and biological models.
Markov chains replace nondeterminism in transition systems with probabilistic uncertainty, and \emph{probabilistic model checking}~\cite{BK08,DBLP:reference/mc/BaierAFK18} provides model checking algorithms.
A key property that probabilistic model checkers answer is: \emph{What is the (precise) probability that a target state is reached (within a
finite number of steps $h$)}? Contrary to classical \emph{qualitative} model
checking and approximate variants of probabilistic model checking,   precise probabilistic model checking must find the total probability
of \emph{all} paths from the initial state to any target state.

Nevertheless, the prevalent ideas in probabilistic model checking are generalizations of qualitative model checking.
Whereas qualitative model checking tracks the states that can reach a target state (or dually, that can be reached from an initial state), probabilistic model checking tracks the $i$-step reachability probability for each state in the chain. The $i{+}1$-step reachability can then be computed via  multiplication with the \emph{transition
matrix}. The scalability concern is that this matrix grows with the state space in the Markov chain.
Mature model checking tools such as \storm~\cite{storm},
Modest~\cite{DBLP:conf/tacas/HartmannsH14}, and \prism{}~\cite{DBLP:conf/cav/KwiatkowskaNP11} utilize a variety of methods to alleviate the state space explosion. Nevertheless various natural models cannot be analyzed by the available
techniques.


In parallel, within the AI community  a different approach to representing a
distribution has emerged, which on first glance can seem unintuitive. Rather than
marginalizing out the paths and tracking reachability probabilities per state, the probabilistic AI community commonly aggregates all \emph{paths} that reach the target
state. At its core, inference is then a weighted sum over all these paths~\cite{chavira2008probabilistic}. This
hinges on the observation that this set of paths
can often be stored more compactly, and that the probability of two paths that share
the same prefix or suffix can be efficiently computed on this concise
representation. This \emph{inference technique} has been used in a variety of
domains in the artificial intelligence (AI) and verification
communities~\cite{HoltzenOOPSLA20,fierens2015inference,DBLP:conf/ijcai/ChakrabortyFMV15,DBLP:conf/ccs/BalutaSSMS19}, but is
not part of any mature probabilistic model checking tools.

This paper theoretically and experimentally compares and contrasts these two approaches. In particular, we describe and motivate
\rubicon{}, a probabilistic model checker that \emph{leverages the successful
probabilistic inference techniques}. We begin with an example 
that explains the core ideas of \rubicon{} followed by the paper structure and key contributions.

\newsavebox{\motivprism}
\begin{lrbox}{\motivprism}
\lstset{language=Prism}   
  \begin{lstlisting}[mathescape=true]
const double $p_1$, $p_2$, $p_3$, $q_1$, $q_2$, $q_3$;
module F1
 $c_1$ :  bool init false;
 [a] !$c_1$ ->  $ p_1$: ($c_1'$=1) +  $ 1{-}p_1$: ($c_1'$=0);
 [a]  $c_1$ ->  $ q_1$: ($c_1'$=0) +  $ 1{-}q_1$: ($c_1'$=1);
endmodule
module F2 = F1[$c_1$=$c_2$,$p_1$=$p_2$,$q_1$=$q_2$]
module F3 = F1[$c_1$=$c_3$,$p_1$=$p_3$,$q_1$=$q_3$]
label "allStrike" = $c_1$ &  $\;c_2$ &  $\;c_3$;
\end{lstlisting}
\end{lrbox}

\begin{figure}[t]
    \centering

    \subfigure[Motivating factory Markov chain with $s_i = \semantics{c_i=0}, t_i = \semantics{c_i=1}$.]{
        \scalebox{0.88}{
\begin{tikzpicture}
	\node[state,initial,initial text=,initial distance=3mm,initial where=above] (s00) {$s_1$};
	\node[state, right=of s00] (s01) {$t_1$};
	
	\node[right=2.1cm of s01] (dots) {\LARGE$\cdots$};
	\node[left=0.3cm of dots] (p0) {\Huge$\vert\vert$};
	\node[right=0.3cm of dots] (p1) {\Huge$\vert\vert$};
	
	\node[state, right=1.1cm of p1,,initial,initial text=,initial distance=3mm,initial where=above] (sn0) {$s_n$};
	\node[state, right=of sn0] (sn1) {$t_n$};
	
	\draw[->] (s00) edge[bend left] node[above] {$p_1$} (s01);
	\draw[->] (s01) edge[bend left] node[above] {$q_1$} (s00);
	\draw[->] (s00) edge[loop left] node[above,yshift=0.3em,align=center] {\tiny{$\bar{p}_1 \colonequals$}\\$1{-}{p}_1$} (s00);
	\draw[->] (s01) edge[loop right] node[above,yshift=0.3em,align=center] {\tiny{$\bar{q}_1 \colonequals$}\\$1{-}{q}_1$} (s01);

	\draw[->] (sn0) edge[bend left] node[above] {$p_n$} (sn1);
	\draw[->] (sn1) edge[bend left] node[above] {$q_n$} (sn0);
	
	\draw[->] (sn0) edge[loop left] node[above,yshift=0.3em,align=center] {\tiny{$\bar{p}_n \colonequals$}\\$1{-}{p}_n$} (sn0);
	\draw[->] (sn1) edge[loop right] node[above,yshift=0.3em,align=center] {\tiny{$\bar{q}_n \colonequals$}\\$1{-}{q}_n$} (sn1);
\end{tikzpicture}
}
\label{fig:motivmodel}
}
\vspace{-3mm}
    \subfigure[A \prism{} model of (a) with 3 factories.]{
    \raisebox{5em}{
      \usebox{\motivprism}
}
      \label{fig:motivprism}
}~
\subfigure[Relative scaling.]{
\begin{tikzpicture}
   \begin{axis}[
		height=3.5cm,
   width=3cm,
		grid=major,
		ticklabel style = {font=\tiny},
   xlabel={\# Parallel Chains},
   ylabel={time in s},
   y label style={at={(axis description cs:.4,.5)},,anchor=south},
   legend style={at={(5em,5em)},anchor=north}
   ]
   \addplot[mark=none, mark = star] table [x index={0}, y index={1}] {\motiv}; \label{plt:rubicon}
   \addplot[mark=none, red, mark = o ] table [x index={0}, y index={2}] {\motiv}; \label{plt:stormexpl}
   \addplot[mark=none, blue, mark = square] table [x index={0}, y index={3}] {\motiv} ;\label{plt:stormsym}
   \addplot[mark=none, green,  thick, mark = diamond] table [x index={0}, y index={4}] {\motiv};\label{plt:prism}
   
 \end{axis}
\end{tikzpicture}
\label{fig:motivperf}
}
\subfigure[BDD]{
 \begin{tikzpicture}
    \def\lvl{24pt}
    \node (s00) at (0bp,0bp) [bddnode] {$c_1^{(1)}$};
    \node (s10) at ($(s00) + (-8bp, -\lvl)$) [bddnode] {$c_2^{(1)}$};
    \node (s20) at ($(s10) + (-8bp, -\lvl)$) [bddnode] {$c_3^{(1)}$};
    \node (t) at ($(s20) + (-8bp, -\lvl)$) [bddterminal] {$T$};
    \node (f) at ($(s00) + (8bp, -3*\lvl)$) [bddterminal] {$F$};
    \begin{scope}[on background layer]
       \draw [thick, ->] (s00) -- (s10) node[midway,above left] {$p_1$};
       \draw [thick, ->] (s10) -- (s20) node[midway,above left] {$p_2$};
       \draw [thick, ->] (s20) -- (t) node[midway,above left] {$p_3$};
       \draw [thick, lowedge, ->] (s00) -- (f) node[midway,above right] {$\bar{p}_1$};
       \draw [thick, lowedge, ->] (s10) -- (f) node[midway,above ] {$\bar{p}_2$};
       \draw [thick, lowedge, ->] (s20) -- (f) node[midway,above] {$\bar{p}_3$};
    \end{scope}

  \end{tikzpicture}
  \label{fig:motiv}
}
\vspace{-1mm}
\caption{Motivating example. Figure~\ref{fig:motivperf} compares the performance
  of \rubicon{} (\ref{plt:rubicon}), \storm{}'s explicit engine
  (\ref{plt:stormexpl}), \storm{}'s symbolic engine (\ref{plt:stormsym}) and
  \prism{} (\ref{plt:prism}) when invoked on a (b) with arbitrarily fixed
  (different) constants for $p_i,q_i$ and horizon $h=10$. Times are in seconds, with a time-out of
  30 minutes.}
\end{figure}
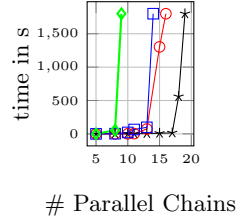
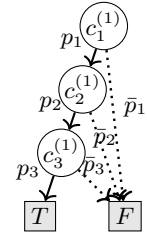

\mysubsubsection{Motivating Example}\label{sec:motivating_example}
Consider the example illustrated in Fig.~\ref{fig:motivmodel}. Suppose there are
$n$ factories. Each day, the workers at each factory collectively decide
whether or not to strike. To simplify, we model each factory ($i$) with two
states, striking ($t_i$) and not striking ($s_i$). Furthermore, since no two
factories are identical, we take the probability to begin striking ($p_i$) and
to stop striking ($q_i$) to be different for each factory. Assuming that each
factory transitions synchronously and in parallel with the others, we query:
``what is the probability that all the factories are simultaneously striking
within $h$ days?''

Despite its simplicity, we observe that state-of-the-art model checkers
like \storm{} and \prism{} do not scale beyond $15$ 
factories.\footnote{Section~\ref{sec:experiments} describes the
  experimental apparatus and our choice of comparisons.} For example, Figure~\ref{fig:motivprism} provides a
\prism{} encoding for this simple model (we show the instance with $3$
factories), where a Boolean variable $c_i$ is used to encode the state of each factory. The ``\texttt{allStrike}'' label identifies the target
state. Figure~\ref{fig:motivperf} shows the run time for an increasing
number of factories. While all methods eventually time out, \rubicon{} scales to systems with an order of magnitude more states.



\emph{Why is this problem hard?}
To understand the issue with scalability, observe that tools such as \storm{} and \prism{}  store the transition matrix, either explicitly or symbolically using algebraic
decision diagrams (ADDs).
Every distinct entry of this transition matrix needs to be represented; in the case of ADDs using a unique leaf node.
Because each factory in our example has a different probability of going on strike, that means each subset of factories will likely have a unique probability of jointly going on strike. Hence, the transition matrix then will have a number of distinct probabilities that is exponential in the number of factories, and its representation as an ADD must blow up in size.
Concretely, for 10 factories,
the size of the ADD representing the transition matrix has 1.9 million
nodes. Moreover, the explicit engine fails due to the dense nature of the underlying transition matrix. We discuss this method in Sec.~\ref{sec:add}.
\emph{How to overcome this limitation?}  
This problematic combinatorial explosion is often unnecessary. For the sake of intuition, consider the simple case where the horizon is 1. Still, the standard transition matrix representations blow up exponentially with the number of factories $n$.
Yet, the probability of reaching the ``\texttt{allStrike}'' state is easy to compute, even when $n$ grows: it is $p_1 \cdot p_2 \cdots p_n$.

\rubicon{} aims to compute probabilities in this compact \emph{factorized} way by
representing the computation as a binary decision diagram (BDD). Fig.~\ref{fig:motiv} gives an example of
such a BDD, for three factories and a horizon of one. A key property of this
BDD, elaborated in \secref{add}, is that it can be interpreted as a
\emph{parametric Markov chain}, where the weight of each edge corresponds with
the probability of a particular factory striking. Then, the probability that the
goal state is reached is given by the weighted sum of paths terminating in $T$:
for this instance, there is a single such path with weight $p_1 \cdot p_2 \cdot
p_3$. These BDDs are tree-like Markov-chains, so model checking can be performed
in time linear in the size of the BDD using dynamic programming. Essentially,
the BDD represents the set of paths that reach a
target state---an idea common in probabilistic inference.

To construct this BDD, we propose to encode our reachability query
symbolically as a \emph{weighted model counting} (WMC) query on a logical formula. By compiling that formula into a BDD, we obtain a diagram where computing the query probability can be done efficiently (in the size of the BDD).
Concretely for Fig.~\ref{fig:motiv}, the BDD represents the formula $c_1^{(1)} \land c_2^{(1)} \land c_3^{(1)}$, which encodes
all paths through the chain that terminate in the goal state (all factories strike on day 1). For this example and this horizon, this is a single path.
WMC is a well-known strategy for probabilistic inference 
and is currently the among the state-of-the-art approaches for discrete graphical
models~\cite{chavira2008probabilistic}, discrete probabilistic
programs~\cite{HoltzenOOPSLA20}, and probabilistic logic
programs~\cite{fierens2015inference}.



In general, the exponential growth of the number of paths might seem like it dooms this approach: for $n=3$ factories and 
horizon $h=1$, we need to only represent $8$ paths, but for $h=2$, we would
need to consider $64$ different paths, and so on. However, a key insight is that,
for many systems -- such as the factory example -- the structural compression of BDDs allows a concise
representation of exponentially many paths, all \emph{while} being parametric
over path probabilities~(see Sec.~\ref{sec:dd}). To see why, observe that in the above
discussion, the state of each factory is
\emph{independent} of the other factories: independence, and its natural generalizations
like \emph{conditional} and \emph{contextual} independence, are the driving force behind many
successful probabilistic inference algorithms~\cite{koller2009probabilistic}. Succinctly, the key advantage of
\rubicon{} is that it exploits a form of structure that has thus far been
under-exploited by model checkers, which is why it scales
to more parallel factories than the existing approaches on the hard task. In Section~\ref{sec:experiments} we consider an 
extension to this motivating example that adds 
dependencies between factories. This dependency (or rather, the accompanying increase in the size of the underlying MC) significantly decreases scalability for
the existing approaches but negligibly affects \rubicon{}.

This leads to the task: \emph{how does one go from a \prism{} model to a concise
BDD efficiently}? To do this, \rubicon{} leverages
a novel translation from \prism{} models into a probabilistic programming
language called \dice{} (outlined in Section~\ref{sec:translation}). 




\mysubsubsection{Contribution and Structure}
\label{sec:contr}
Inspired by the example, we contribute conceptual and empirical
arguments for leveraging BDD-based probabilistic inference in model checking.
Concretely:
\begin{compactenum}
 \item We demonstrate fundamental advantages in using probabilistic inference on a natural class of models (Sec.~\ref{sec:motivating_example} and \ref{sec:experiments}).
\item We explain these advantages by showing the fundamental differences between existing
  model checking approaches and probabilistic inference (Sec.~\ref{sec:add}
  and \ref{sec:dd}). 
To that end, Section~\ref{sec:dd} presents probabilistic inference based on an operational and a logical perspective and combines these perspectives. 
\item We leverage those insights to build \rubicon{}, a tool that transpiles \prism{} to \dice{}, a
  probabilistic programming language
  (Sec.~\ref{sec:translation}).
\item We demonstrate that \rubicon{} indeed attains an 
order-of-magnitude scaling improvement on several natural problems
including sampling from parametric Markov chains and verifying network
protocol stabilization (Sec.~\ref{sec:experiments}).
\end{compactenum}
Ultimately we argue that \rubicon{} makes a valuable contribution to the
portfolio of probabilistic model checking backends, and brings to bear the
extensive developments on probabilistic inference to
well-known model checking problems.

\section{Preliminaries and Problem Statement}
\label{sec:probstatement}
We state the problem formally and recap relevant concepts. See~\cite{BK08} for details. We sometimes use $\bar{p}$ to denote $1 - p$.
A \emph{Markov chain} (MC) is a tuple $\mc = \langle S, \iota, P, T \rangle$ with $S$ a (finite) set of \emph{states}, $\iota \in S$ the \emph{initial state}, $P\colon S \rightarrow \Distr(S)$ the \emph{transition function}, and $T$ a set of \emph{target states} $T \subseteq S$, where $\Distr(S)$ is the set of distributions over a (finite) set $S$.
We  write $P(s,s')$ to denote $P(s)(s')$ and  call $P$ a \emph{transition matrix}. 
 The successors of $s$ are $\Succ(s) = \{s' \mid P(s,s') > 0 \}$. 
 To support MCs with billions of states, we may describe MCs symbolically, e.g., with \prism~\cite{DBLP:conf/cav/KwiatkowskaNP11} or as a probabilistic program~\cite{DBLP:journals/jcss/Kozen81,DBLP:conf/birthday/KatoenGJKO15}. 
For such a symbolic description  $\prog$, we denote the corresponding MC with $\semantics{\prog}$. States then  reflect assignments to symbolic variables.

A \emph{path} $\pi = s_0 \hdots s_n$ is a sequence of states, $\pi \in S^{+}$. We use $\last{\pi}$ to denote the \emph{last state} $s_n$, and the \emph{length} of $\pi$ above is $n$ and is denoted $|\pi|$. 
Let $\hPaths$ denote the paths of length $h$. 
The probability of a path is the product of the transition probabilities, and may be defined inductively by $\Pr(s) = 1$, $\Pr(\pi \cdot s) = \Pr(\pi) \cdot P(\last{\pi},s)$. 
For a fixed \emph{horizon} $h$ and set of states $T$, let the set
$\seteventuallybfrom{s}{h}{T} = \{ \pi \mid \pi_0 = s \land |\pi| \leq h \land  \last{\pi} \in T
\land \forall i < |\pi|.~\pi_i \not\in T\}$ denote paths from $s$ of length at most $h$
that terminate at a state contained in $T$.
Furthermore, let $\Pr_\mc(s \models \eventuallyb{h} T) = \sum_{\pi \in
  \seteventuallybfrom{s}{h}{T}} \Pr(\pi)$ describe the probability to reach $T$ within $h$ steps. We simplify notation when $s = \iota$ and write $\seteventuallyb{h}{T}$ and $\Pr_\mc(\eventuallyb{h} T)$, respectively. We omit $\mc$ whenever that is clear from the context.

\begin{figure}[t]
\centering
\subfigure[Toy-example $\mc$]{
\begin{tikzpicture}[ every node/.style={font=\scriptsize}]
		\node[initial,initial text=,sstate] (s1) {$0{,}0$};
		\node[sstate,below=0.8cm of s1] (s2) {$0{,}1$};
		\node[sstate,right=1.2cm of s1,accepting] (s3) {$1{,}0$};
		\node[sstate,right=1.2cm of s2] (s4) {$1{,}1$};
		
		\draw[->] (s1) -- node[left] {$0.4$} (s2);
		\draw[->] (s3) edge[bend left] node[right] {$0.4$} (s4);
		\draw[->] (s3) -- node[above] {$0.6$} (s1);
		
		\draw[->] (s1) edge[loop above] node[left] {$0.6$} (s1);
		\draw[->] (s2) -- node[left] {$0.5$} (s3);
		\draw[->] (s2) -- node[above] {$0.5$} (s4);
		\draw[->] (s4) edge[bend left] node[right] {$0.5$} (s3);
		\draw[->] (s4) edge[loop right] node[right] {$0.5$} (s4);
		
	\end{tikzpicture}
	\label{fig:runningmini}
}
\subfigure[pMC $\mc'$]{
\begin{tikzpicture}[ every node/.style={font=\scriptsize}]
		\node[initial,initial text=,sstate] (s1) {$0{,}0$};
		\node[sstate,below=0.8cm of s1] (s2) {$0{,}1$};
		\node[sstate,right=1.2cm of s1,accepting] (s3) {$1{,}0$};
		\node[sstate,right=1.2cm of s2] (s4) {$1{,}1$};
		
		\draw[->] (s1) -- node[left] {$p$} (s2);
		\draw[->] (s3) edge[bend left] node[right] {$p$} (s4);
		\draw[->] (s3) -- node[above] {$1{-}p$} (s1);
		
		\draw[->] (s1) edge[loop above] node[left] {$1{-}p$} (s1);
		\draw[->] (s2) -- node[left] {$q$} (s3);
		\draw[->] (s2) -- node[above] {$1{-}q$} (s4);
		\draw[->] (s4) edge[bend left] node[right] {$q$} (s3);
		\draw[->] (s4) edge node[right] {$1{-}q$} (s1);
		
	\end{tikzpicture}
	\label{fig:runningminialt}
	\label{fig:pmc}
}
\subfigure[For $\mc$: $P$ as  ADD]{
\begin{tikzpicture}
      
    \def\lvl{14pt}
    \node (s0) at (0bp,0bp) [bddnode] {$y$};

    \node (n0) at ($(s0) + (-30bp, -0.6*\lvl)$) [bddnode] {$y'$};
    
    \node (n01) at ($(n0) + (20bp, -0.6*\lvl)$) [bddnode] {$x$};
    
     \node (n010) at ($(n01) + (-15bp, -\lvl)$) [bddnode] {$x'$};
    \node (n011) at ($(n01) + (15bp, -\lvl)$) [bddnode] {$x'$};
    \node (n100) at ($(n01) + (+40bp, -\lvl)$) [bddnode] {$x'$};
    \node (n000) at ($(n01) + (-40bp, -\lvl)$) [bddnode] {$x'$};

    \node (zero1) at ($(n010) + (0, -1.5*\lvl)$) [bddterminal] {$0$};
    \node (t0p) at ($(n000) + (0, -1.5*\lvl)$) [bddterminal] {$0.6$};
    \node (t1p) at ($(n011) + (0, -1.5*\lvl)$) [bddterminal] {$0.4$};
    \node (half) at ($(n100) + (0, -1.5*\lvl)$) [bddterminal] {$0.5$};
   
    \begin{scope}[on background layer]
       \draw [lowedge] (s0) -- (n0);
       \draw [highedge] (s0) -- (n100);

       \draw [lowedge] (n0) -- (n000);
       \draw [highedge] (n0) -- (n01);

       \draw [lowedge] (n01) -- (n010);
       \draw [highedge] (n01) -- (n011);
       
       \draw [lowedge] (n000) -- (t0p);
       \draw [highedge] (n000) -- (zero1);
        \draw [lowedge] (n100) -- (zero1);
       \draw [highedge] (n100) -- (half);
        \draw [lowedge] (n010) -- (t1p);
       \draw [highedge] (n010) -- (zero1);
        \draw [lowedge] (n011) -- (zero1);
       \draw [highedge] (n011) -- (t1p);
    \end{scope}

  \end{tikzpicture}
  \label{fig:exadd}
}
\caption{(a) MC toy example (b) (distinct) pMC toy example (c) ADD transition
  matrix }
\end{figure}
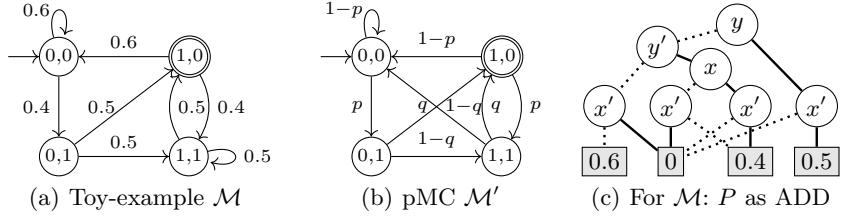 
\begin{mdframed}
\vspace{-1mm}
\textbf{Formal Problem:}
Given an MC $\mc$ and a horizon $h$, compute $\Pr_\mc(\eventuallyb{h} T)$.
\end{mdframed}
\begin{example}
\label{ex:mc}
For conciseness, we introduce a toy example MC $\mc$ in
Fig.~\ref{fig:runningmini}. For horizon $h=3$, there are three paths that reach
state $\langle 1{,}0 \rangle$: For example the path $\langle 0,0 \rangle \langle
0,1 \rangle \langle 1,0 \rangle$ with corresponding reachability probability $0.4
\cdot 0.5$. The reachability probability
$\Pr_\mc(\eventuallyb{3} \{ \langle 1,0 \rangle \}) = 0.42$.  
\end{example}

It is helpful to separate the topology and the probabilities.
We do this by means of a \emph{parametric MC} (pMC)~\cite{DBLP:conf/ictac/Daws04}. 
A pMC over a fixed set of parameters $\vec{p}$ generalises MCs by allowing for a transition function that maps to $\QQ[\vec{p}]$, i.e., to polynomials over these variables~\cite{DBLP:conf/ictac/Daws04}. 
 A pMC and a \emph{valuation} of parameters $\vec{u}\colon \vec{p} \rightarrow \RR$  describe a MC by replacing $\vec{p}$ with $\vec{u}$ in the transition function $P$ to obtain $P[\vec{u}]$. If $P[\vec{u}](s)$ is a distribution for every $s$, then we call $\vec{u}$ a \emph{well-defined} valuation. We can then think about a pMC $\mc$ as a generator of a set of MCs $\{ \mc[\vec{u}] \mid \vec{u} \text{ well-defined} \}$. Fig.~\ref{fig:pmc} shows a pMC; any valuation $\vec{u}$ with $\vec{u}(p), \vec{u}(q) \in [0,1]$ is well-defined.
 We consider the following associated problem:
\begin{mdframed}
\vspace{-1mm}
\textbf{Parameter Sampling:}
Given a pMC $\mc$, a finite set of well-defined valuations $U$, and a horizon $h$, compute $\Pr_{\mc[\vec{u}]}(\eventuallyb{h} T)$ for each $\vec{u} \in U$.
\vspace{-1mm}
\end{mdframed}

We recap  binary \emph{decision
diagrams} (BDDs) and their generalization into algebraic decision diagrams (ADDs,
a.k.a. multi-terminal BDDs). ADDs over a set of
variables $X$ are directed acyclic graphs whose vertices $V$ can
be partitioned into \emph{terminal nodes} $V_t$ without successors and \emph{inner
  nodes} $V_i$ with two successors. Each terminal node is labeled with a polynomial over some parameters $\vec{p}$ (or just to constants in $\QQ$),
$\valmap\colon V_t \rightarrow \QQ[\vec{p}]$, and each inner node $V_i$ with a
variable, $\varmap\colon V_i \rightarrow X$. One node is the root node $v_0$.
Edges are described by the two successor functions $E_0\colon V_i \rightarrow V$
and $E_1\colon V_i \rightarrow V$. A BDD is an ADD with exactly two terminals labeled
$T$ and $F$.
Formally, we denote an ADD by the tuple $\langle V, v_0, X, \varmap, \valmap, E_0, E_1 \rangle$.
ADDs describe functions $f\colon \BB^{X} \rightarrow \QQ[\vec{p}]$ (described by a path in the underlying graph and the label of the corresponding terminal node). As finite sets can be encoded with bit vectors, ADDs represent functions from (tuples of) finite sets to polynomials.

\begin{example}
The transition matrix $P$ of the MC in Fig.~\ref{fig:runningmini} maps states, encoded by bit vectors, $\langle x, y \rangle, \langle x', y' \rangle$ to the probabilities to move from state $\langle x, y \rangle$ to $\langle x', y' \rangle$. Fig.~\ref{fig:exadd} shows the corresponding ADD.\footnote{The ADD also depends on the variable order, which we assume fixed for conciseness.}
\end{example}


\section{A Model Checking Perspective}
\label{sec:add}
We briefly analyze the de-facto standard approach to symbolic probabilistic model checking of finite-horizon reachability probabilities.
It is an
adaptation of qualitative model checking, in which we track the
(backward) reachable states. This set can be thought of as a mapping from
states to a Boolean indicating whether a target state can be reached.
 We generalize the mapping to a function that maps every state $s$ to the
probability that we reach $T$ within $i$ steps, denoted $\Pr_\mc(s \models \eventuallyb{i} T)$.
First, it is convenient to construct a transition relation in which the target
states have been made absorbing, i.e., we define a matrix with  $A(s,s') =
P(s,s')$ if $s \not\in T$ and $A(s,s') = [s = s']$\footnote{Where $[x]{=}1$ if $x$ holds and $0$ otherwise.} otherwise.
 The following \emph{Bellman equations} characterize that aforementioned  mapping: 
\begin{align*} \Pr_\mc\big(s \models \eventuallyb{0} T\big) &= [s \in T],\\
  \Pr_\mc\big(s \models \eventuallyb{i} T\big) &= \sum_{s' \in \Succ(s)} \!A(s,s') \cdot \Pr_\mc(s' \models \eventuallyb{i-1} T) \qquad\text{ with $i>0$}.\end{align*} 
The main aspect model checkers take from these equations is that to compute the $h$-step reachability from state $s$, one only needs to combine the $h{-}1$-step reachability from any state $s'$ \emph{and} the transition probabilities $P(s,s')$.  We define a vector $\vec{T}$ with $\vec{T}(s) = [s \in T]$.  The algorithm then iteratively computes and stores the $i$ step reachability for $i = 0$ to $i = h$, e.g. by computing $A^3\cdot \vec{T}$ using $A \cdot (A\cdot (A \cdot \vec{T}))$.
This reasoning is thus \emph{inherently backwards} and \emph{implicitly marginalizing out paths}. In particular, rather than storing the $i$-step paths that lead to the target, one only stores a vector $\vec{x} = A^i \cdot \vec{T}$ that stores for every state $s$ the sum over all $i$-long paths from $s$. 

Explicit representations of matrix $A$ and vector $\vec{x}$ require 
memory at least in the order $|S|$.\footnote{Excluding e.g., partial exploration or sampling which typically are not exact.} To overcome this limitation,
\emph{symbolic} probabilistic model checking stores both $A$ and $A^i \cdot \vec{T}$ as an ADD by considering the matrix as a function from a tuple $\langle s,s' \rangle$ to $A(s,s')$, and $\vec{x}$ as a function from $s$ to $\vec{x}(s)$~\cite{DBLP:conf/tacas/AlfaroKNPS00}.

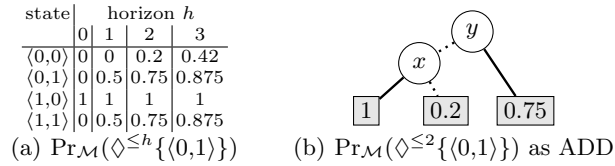
\begin{figure}[t]
\centering
\subfigure[$\Pr_\mc(\eventuallyb{h} \{\langle 0{,}1\rangle\})$]{
\raisebox{20pt}{
\scriptsize{
\begin{tabular}{l|c|c|c|c}
state    & \multicolumn{4}{c}{horizon $h$} \\
        & 0 & 1   & 2          & 3 \\ \hline
$\langle 0{,}0 \rangle $ & 0 & 0   & $0.2$       & $0.42$   \\
$\langle 0{,}1 \rangle $ & 0 & $0.5$     & $0.75$ & $0.875$   \\
$\langle 1{,}0\rangle$ & 1 & 1   & 1          & 1  \\
$\langle 1{,}1\rangle$ & 0 & $0.5$   &  $0.75$         & $0.875$ 
\end{tabular}
}
}
\label{fig:exboundedreach}
}	
\quad
\subfigure[$\Pr_\mc(\eventuallyb{2} \{\langle 0{,}1\rangle\})$ as ADD]{
\begin{tikzpicture}
      
    \def\lvl{12pt}
    \node (s0) at (0bp,0bp) [bddnode] {$y$};

    \node (n0) at ($(s0) + (-20bp, -\lvl)$) [bddnode] {$x$};
    \node (zero1) at ($(n0) + (-20bp, -1.5*\lvl)$) [bddterminal] {$1$};
    \node (t1p) at ($(n0) + (+10bp, -1.5*\lvl)$) [bddterminal] {$0.2$};
    \node (half) at ($(n0) + (40bp, -1.5*\lvl)$) [bddterminal] {$0.75$};
     \node[right=0.6cm of half] (dummy) {};
       \node[left=0.6cm of zero1] (dummy) {};
    \begin{scope}[on background layer]
       \draw [lowedge] (s0) -- (n0);
       \draw [highedge] (s0) -- (half);

       \draw [lowedge] (n0) -- (t1p);
       \draw [highedge] (n0) -- (zero1);

    \end{scope}

\end{tikzpicture}
\label{fig:addres}
}
\caption{Bounded reachability and symbolic model checking for the MC  $\mc$ in Fig.~\ref{fig:runningmini}.}
\label{fig:runningrecex}
\end{figure}
\begin{example}
Reconsider the MC in Fig.~\ref{fig:runningmini}. The $h$-bounded reachability probability $\Pr_\mc(\eventuallyb{h} \{\langle 1,0 \rangle \})$ can be computed as reflected in~Fig.~\ref{fig:exboundedreach}.
The ADD for $P$ is shown in Fig.~\ref{fig:exadd}.  
The ADD for $\vec{x}$ when $h=2$ is shown in Fig.~\ref{fig:addres}. \end{example}

The performance of symbolic probabilistic model checking is  directly
governed by the sizes of these two ADDs. The size of an ADD is
bounded from below by the number of leafs. In qualitative model checking, both ADDs are in fact BDDs, with two leafs.
However, for the ADD representing $A$, this lower bound is given by the number
of different probabilities in the transition matrix.
In the running
example, we have seen that a small program $\prog$ may have an underlying MC
$\semantics{\prog}$ with an exponential state space $S$ and equally many
different transition probabilities.
Symbolic probabilistic model checking also scales badly on some models where $A$
has a concise encoding but $\vec{x}$ has too many different
entries.\footnote{For an interesting example of this, see the ``Queue'' example
  in Section~\ref{sec:experiments}.}
Therefore, model checkers may store $\vec{x}$ partially explicit~\cite{DBLP:conf/tacas/KwiatkowskaNP02}.

The insights above are not new. 
Symbolic probabilistic model checking has advanced~\cite{DBLP:journals/sttt/KleinBCDDKMM18} to create small representations of both $A$ and $\vec{x}$.
In competitions, \storm{} often applies a bisimulation-to-explicit method that extracts an explicit representation of the bisimulation quotient~\cite{DBLP:journals/sttt/DijkP18,storm}. 
Finally, game-based abstraction~\cite{DBLP:journals/fmsd/KattenbeltKNP10,DBLP:conf/tacas/HahnHWZ10} can be seen as a predicate abstraction technique on the ADD level.
However, these methods do not change the computation of the finite horizon reachability probabilities and thus do not overcome the inherent weaknesses of the iterative approach in combination with an ADD-based representation.

\section{A Probabilistic Inference Perspective}
We present four key insights into probabilistic inference.
\textbf{(1)} Section~\ref{sec:operational} shows how probabilistic inference
takes the classical definition as summing over the set of paths, and turns this
definition into an algorithm. In particular, these paths may be stored in a
computation tree.
\textbf{(2)} Section~\ref{sec:logical} gives the traditional reduction from probabilistic
inference to the classical weighted model counting ($\WMC$)
problem~\cite{chavira2008probabilistic,sang2005performing}.
\textbf{(3)}
Section~\ref{sec:wmc} connects this reduction to point (1) by showing that a BDD that represents
this $\WMC$ is \emph{bisimilar} to the computation tree assuming that
the out-degree of every state in the MC is two.
\textbf{(4)} Section~\ref{sec:causalitys} describes and compares the computational
benefits of the BDD representation. In particular, we clarify that enforcing an  out-degree of two is a key ingredient to overcoming one of
the weaknesses of symbolic probabilistic model checking: the number of different probabilities in the underlying MC. 

\label{sec:dd}
\subsection{Operational perspective}
\label{sec:operational}
The following perspective frames (an aspect of) probabilistic inference as a
model transformation.
By definition, the set of all paths -- each
annotated with the transition probabilities -- suffices to extract the reachability
probability. 
These sets of paths may be represented in the computation tree (which is
itself an MC). 
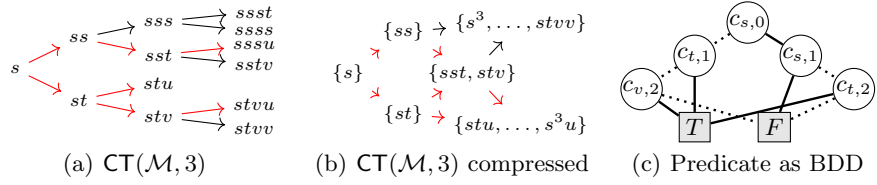
\begin{figure}[t]
%
%
%
\subfigure[$\ct{\mc}{3}$]{
	\def\lvlbct{9pt}
	\def\lvlct{30bp}
\begin{tikzpicture}[ every node/.style={font=\scriptsize}]
	\node[] (s) {$s$};
	\node[] (st) at ($(s) + (0.8*\lvlct,-1.3*\lvlbct)$) {$st$};
	\node[] (ss) at ($(s) + (0.8*\lvlct,1.3*\lvlbct)$) {$ss$};
	\node[] (stu) at ($(st) + (\lvlct,-0.7*\lvlbct)$) {$stv$};
	\node[] (stv) at ($(st) + (\lvlct,0.7*\lvlbct)$) {$stu$};
	\node[] (sss) at ($(ss) + (\lvlct,0.7*\lvlbct)$) {$sss$};
	\node[] (sst) at ($(ss) + (\lvlct,-0.7*\lvlbct)$) {$sst$};
	\node[] (stuu) at ($(stu) + (1.2*\lvlct,-0.5*\lvlbct)$) {$stvv$};
	\node[] (stuv) at ($(stu) + (1.2*\lvlct,0.5*\lvlbct)$) {$stvu$};
	\node[] (ssss) at ($(sss) + (1.2*\lvlct,-0.4*\lvlbct)$) {$ssss$};
	\node[] (ssst) at ($(sss) + (1.2*\lvlct,0.4*\lvlbct)$) {$ssst$};
	\node[] (sstu) at ($(sst) + (1.2*\lvlct,-0.4*\lvlbct)$) {$sstv$};
	\node[] (sstv) at ($(sst) + (1.2*\lvlct,0.4*\lvlbct)$) {$sssu$};
	\draw[->,red] (s) -- node[auto] {} (st);
	\draw[->,red] (s) -- node[auto] {} (ss);
	\draw[->,red] (ss) -- node[auto] {} (sst);
	\draw[->] (ss) -- node[auto] {} (sss);
	\draw[->,red] (st) -- node[auto] {} (stu);
	\draw[->,red] (st) -- node[auto] {} (stv);
	\draw[->] (sss) -- node[auto] {} (ssss);
	\draw[->] (sss) -- node[auto] {} (ssst);
	\draw[->] (sst) -- node[auto] {} (sstu);
	\draw[->,red] (sst) -- node[auto] {} (sstv);
	
	\draw[->,red] (stu) -- node[auto] {} (stuv);
	\draw[->] (stu) -- node[auto] {} (stuu);
\end{tikzpicture}
\label{fig:fullct}
}	
\subfigure[$\ct{\mc}{3}$ compressed]{
	\def\lvlbct{12pt}
	\def\lvlct{26bp}
\begin{tikzpicture}[ every node/.style={font=\scriptsize}]
	\node[] (s) {$\{ s \}$};
	\node[] (st) at ($(s) + (0.8*\lvlct,-1.3*\lvlbct)$) {$\{ st \}$};
	\node[] (ss) at ($(s) + (0.8*\lvlct,1.3*\lvlbct)$) {$\{ ss \}$};
	\node[] (stu) at ($(st) + (\lvlct,1.3*\lvlbct)$) {$\{ sst, stv \}$};
	\node[] (stuv) at ($(stu) + (0.7*\lvlct,-1.6*\lvlbct)$) {$\{ stu, \hdots, s^3u \}$};
	\node[] (sstu) at ($(stu) + (0.7*\lvlct,1.6*\lvlbct)$) {$\{ s^3, \hdots, stvv\}$};
	\draw[->,red] (s) -- node[auto] {} (st);
	\draw[->,red] (s) -- node[auto] {} (ss);
	\draw[->] (ss) -- node[auto] {} (sstu);
	\draw[->,red] (ss) -- node[auto] {} (stu);
	\draw[->,red] (st) -- node[auto] {} (stuv);
	\draw[->,red] (st) -- node[auto] {} (stu);
	\draw[->] (stu) -- node[auto] {} (sstu);	
	\draw[->,red] (stu) -- node[auto] {} (stuv);
\end{tikzpicture}
\label{fig:compressedct}
}	
\subfigure[Predicate as BDD]{
\begin{tikzpicture}
	
\def\lvl{18pt}
    \node (s0) at (0bp,0bp) [bddnode] {$c_{s,0}$};
	\node (s1) at ($(s0) + (20bp, -0.7*\lvl)$) [bddnode] {$c_{s,1}$};
	\node (t1) at ($(s0) + (-20bp, -0.7*\lvl)$) [bddnode] {$c_{t,1}$};
	\node (t2) at ($(s1) + (20bp, -0.7*\lvl)$) [bddnode] {$c_{t,2}$};
	\node (u2) at ($(t1) + (-20bp, -0.7*\lvl)$) [bddnode] {$c_{v,2}$};

    \node (zero1) at ($(s0) + (-20bp, -2.2*\lvl)$) [bddterminal] {$T$};
    \node (t1p) at ($(s0) + (+10bp, -2.2*\lvl)$) [bddterminal] {$F$};
    \begin{scope}[on background layer]
        \draw [highedge] (s0) -- (s1);
       \draw [lowedge] (s0) -- (t1);
         \draw [highedge] (s1) -- (t1p);
       \draw [lowedge] (s1) -- (t2);
       \draw [highedge] (t2) -- (zero1);
       \draw [lowedge] (t2) -- (t1p);
       
       \draw [lowedge] (t1) -- (u2);
       \draw [highedge] (t1) -- (zero1);

       \draw [lowedge] (u2) -- (t1p);
       \draw [highedge] (u2) -- (zero1);
    \end{scope}
    \end{tikzpicture}
\label{fig:ctbdd}
}
\caption{The computation tree for $\mc$ and horizon $3$ and its compression.  We
  label states as $s{=}\langle 0{,}0\rangle$, $t{=}\langle 0{,}1 \rangle$, $u{=}\langle 1{,}0\rangle,v{=}\langle 1{,}1 \rangle$. Probabilities are omitted for conciseness.}
\label{fig:cts}
\end{figure}

\begin{example}
	We continue from Ex.~\ref{ex:mc}. 
	We put all paths of length three in a computation tree in Fig.~\ref{fig:fullct} (cf.\ the caption for state identifiers).
	The three paths that reach the target are highlighted in red. The MC is highly redundant.
We may compress to the MC in Fig.~\ref{fig:compressedct}. 
\end{example}
\begin{definition}
For MC $\mc$ and horizon $h$, the computation tree (CT) $\ct{\mc}{h} =
\langle \hPaths, \iota, P', T' \rangle$ is an MC 
with states corresponding to paths in $\mc$, i.e., $\hPaths^\mc$, initial
state $\iota$, 
target states 
 $T'= \seteventuallyb{h}{T}$, and transition relation
\begin{equation}
  P'(\pi, \pi') = \begin{cases}
    P(\last{\pi}, s) & \text{if } \last{\pi} \notin T \wedge \pi' = \pi.s,\\ 
    [\last{\pi} \in T \wedge \pi' = \pi ] & \text{otherwise}.
 \end{cases}
 \end{equation}
\end{definition}
The CT contains (up to renaming) the same paths to the target as the original MC. Notice that after $h$ transitions, all paths are in a sink state, and thus we can drop the step bound from the property and consider either finite or indefinite horizons. The latter considers all paths that eventually reach the target. We denote the probability mass of these paths with $\Pr_\mc(s \models \eventually T)$ and refer to \cite{BK08} for formal details.\footnote{Alternatively, on acyclic models, a large step bound $h > |S|$ suffices.}
Then, we may compute bounded reachability probabilities in the
original MC by analysing unbounded reachability in the CT: 
	\begin{center} $\Pr_\mc(\eventuallyb{h} T) = \Pr_{\ct{\mc}{h}}(\eventuallyb{h} T') = \Pr_{\ct{\mc}{h}}(\eventually T').$\end{center}
The nodes in the CT have a natural topological ordering. The  unbounded reachability probability is then  computed (efficiently in CT's size) using dynamic
programming (i.e., topological value iteration) on the Bellman equation for $s
\not\in T$:
\begin{center}
  $\Pr_\mc(s \models \eventually T) = \sum_{s' \in \Succ(s)} P(s,s') \cdot \Pr_\mc(s' \models \eventually T).$
\end{center}
For pMCs, the right-hand side naturally is a factorised form of the  \emph{solution function} $f$ that maps parameter values to the induced reachability probability, i.e. $f(\vec{u}) = \Pr_{\mc[\vec{u]}}(\eventuallyb{h}T)$~\cite{DBLP:conf/ictac/Daws04,DBLP:journals/sttt/HahnHZ11,DBLP:conf/cav/DehnertJJCVBKA15}. For bounded reachability (or acyclic pMCs), this function amounts to a sum over all paths with every path reflected by a term of a polynomial, i.e., the sum is a polynomial. In sum-of-terms representation, the polynomial can be exponential in the number of parameters~\cite{DBLP:journals/iandc/BaierHHJKK20}. 

For computational efficiency, we need a smaller representation of the CT. 
As we only consider reachability of $T$, we may simplify~\cite{DBLP:conf/tacas/KatoenKZJ07} the notion of (weak) bisimulation~\cite{DBLP:conf/cav/BaierH97} (in the formulation of~\cite{DBLP:conf/concur/JansenGTY20}) to the following definition. 
\begin{definition} 
For $\mc$ with states $S$, a relation $\mathcal{R} \subseteq S \times S$ is a (weak) bisimulation (with respect to $T$) if $s \mathcal{R} s'$ implies $\Pr_{\mc}(s \models \eventually T) = \Pr_{\mc}(s' \models \eventually T)$.  
	Two states $s, s'$ are (weakly) bisimilar (with respect to $T$) if 
	$\Pr_{\mc}(s \models \eventually T) = \Pr_{\mc}(s' \models \eventually T)$
	\end{definition}
Two MCs $\mc, \mc'$ are bisimilar, denoted $\mc \bisim \mc'$ if the initial states are bisimilar in the disjoint union of the MCs. 
It holds by definition that if $\mc \bisim \mc'$, then $\Pr_\mc(\eventually T) = \Pr_{\mc'}(\eventually T')$. The notion of bisimulation can be lifted to pMCs~\cite{DBLP:journals/sttt/HahnHZ11}. 
 \begin{mdframed}
\textbf{Idea 1:} Given a symbolic description $\prog$ of a MC $\semantics{\prog}$, efficiently construct a concise MC $\mc$ that is bisimilar to $\ct{\semantics{\prog}}{h}$.
\end{mdframed}
Indeed, the (compressed) CT
in Fig.~\ref{fig:compressedct} and 
Fig.~\ref{fig:fullct} are bisimilar. 
We remark that we do not necessarily compute the bisimulation
quotient of $\ct{\semantics{\prog}}{h}$.


\subsection{Logical perspective}
\label{sec:logical}

The previous section defined weakly bisimilar chains and showed computational
advantages, but did not present an algorithm. In this section we frame the
finite horizon reachability probability as a logical query known as \emph{weighted
model counting $(\WMC)$}. In the next section we will show how this logical
perspective yields an algorithm for constructing bisimilar MCs.

Weighted model counting is well-known as an effective reduction for probabilistic inference~\cite{sang2005performing,chavira2008probabilistic}. Let $\varphi$ be a
logical sentence over variables $C$. The \emph{weight function}
$W_C \colon C \rightarrow \RRnn$ assigns a weight to each
logical variable. A \emph{total variable assignment} $\eta \colon C \rightarrow
\{ 0, 1 \}$ by definition has weight $\weight(\eta) = \prod_{c \in C} W_C(c)
\eta(c) + (1-W_C(c)) \cdot (1-\eta(c))$.
Then the \emph{weighted model count} for $\varphi$ given $W$ is
$\WMC(\varphi,W_C) = \sum_{\eta \models \varphi} \weight(\eta)$. Formally, we
desire to compute a reachability query using a $\WMC$ query in the following sense:
 \begin{mdframed}
 \textbf{Idea 2:} Given an MC
$\mc$, efficiently construct a predicate $\varphi_{\mc,
  h}^C$ and a weight-function $W_C$ such that
$\Pr_\mc(\eventuallyb{h} T) = \WMC(\varphi_{\mc,h}^\coins, W_C)$.
 \end{mdframed}
Consider initially the simplified case when the MC $\mc$ is \emph{binary}: every
state has at most two successors. In this case producing $(\varphi_{\mc,
  h}^\coins, W_C)$ is straightforward:

\begin{example}
\label{ex:pathspredicate}
Consider the MC in Fig.~\ref{fig:runningmini}, and note that it is binary. We introduce logical variables called
\emph{state/step coins} $\coins = \{ \coin_{s,i} \mid s \in S, i < h \}$ for
every state and step. Assignments to these coins denote choices of transitions
at particular times: if the chain is in state $s$ at step $i$, then it takes the transition to the lexicographically first successor of $s$ if $\coin_{s,i}$ is true and otherwise takes the transition to the lexicographically second successor.
To construct the predicate $\varphi_{\mc, 3}^\coins$, we will need to write a logical sentence on coins whose
models encode accepting paths (red paths) in the CT in Fig.~\ref{fig:fullct}.

We start in state $s=\langle 0,0 \rangle$ (using state labels from the caption
of Fig.~\ref{fig:cts}).
We order states as $s = \langle 0,0 \rangle < t = \langle 0,1 \rangle <  u = \langle 1, 0
\rangle < v = \langle 1,1 \rangle$.
Then, $c_{s,0}$ is true if the chain
transitions into state $s$ at time 0 and false if it transitions to
state $t$ at time 0. So, one path from $s$ to the target node $\langle
1,0 \rangle$ is given by the logical sentence $(c_{s,0} \land \neg c_{s,1}
\land c_{t,2})$. The full
predicate $\varphi^C_{\mc,3}$ is therefore:
\[  \varphi^C_{\mc,3} =
  (c_{s,0} \land \neg c_{s,1} \land  c_{t,2}) \lor
  (\neg c_{s,0} \land c_{t, 1} ) \lor
  (\neg c_{s,0} \land \neg c_{t, 1} \land c_{v, 2}).
\]
Each model of this sentence is a single path to the target.
This predicate $\varphi^C_{\mc,h}$
can clearly be constructed by considering all possible paths through the chain,
but later on we will show how to build it more efficiently.

Finally, we fix $W_C$: The weight for each coin is directly given by the transition probability to the lexicographically first successor:
for $0 \leq i < h$, 
$W_C(c_{s,i}) = 0.6$ and $W_C(c_{t,i}) = W_C(c_{v,i}) = 0.5$. The
$\WMC$ is indeed $0.42$, reflecting Ex.~\ref{ex:mc}. 
\end{example}

When the MC is not binary, it suffices to
limit the out-degree of an MC to be at most two by adding
auxiliary states, hence binarizing all transitions, cf.~Appendix~\ref{sec:binary}.

\subsection{Connecting the Operational and the Logical Perspective}
\label{sec:wmc}
Now that we have reduced bounded reachability to weighted model counting, we
reach a natural question: how do we perform $\WMC$?\footnote{In this paper, we concentrate on
  reductions to \emph{exact} $\WMC$, leaving approximate
  approaches for future work~\cite{DBLP:conf/ijcai/ChakrabortyFMV15}.}
Various approaches to performing $\WMC$ have been explored; a prominent approach is
to compile the logical function into a binary decision diagram (BDD), which
supports fast weighted model counting~\cite{darwiche2002knowledge}. In this
paper, we investigate the use of a BDD-driven approach for two reasons:
\begin{enumerate*}[label=(\roman*)]
\item BDDs admit straightforward support for
parametric models. 
\item BDDs provide a direct
connection between the logical and operational perspectives.
\end{enumerate*}
To start, observe that the graph of the BDD, together with the
weights, can be interpreted as an MC: 
\begin{definition}
	Let $\varphi^X$ be a propositional formula over variables $X$ and ${<}_X$ an ordering on $X$. Let $\bdd{\varphi^X}{{<}_X} = \langle V, v_0, X, \varmap, \valmap, E_0, E_1 \rangle$ be the corresponding BDD, and let $W$ be a weight function on $X$ with $0 \leq W(x) \leq 1$.
	We define the MC $\bddw{\varphi^X}{{<}_X}{W} = \langle S, \iota, P, T \rangle$ with $S = V$, $\iota = v_0$, $P(s) = \{ E_0(s) \mapsto W(\varmap(s)), E_1(s) \mapsto 1-W(\varmap(s)) \}$ and $T = \{ v \in V \mid \valmap(v) = 1 \}$.
\end{definition}
These BDDs are intimately related to the computation trees
discussed before. 
For a binary MC $\mc$, the tree $\ct{\mc}{h}$ is binary and can
be considered as a (not necessarily reduced) BDD. 
More formally, let us construct $\bddw{\varphi_{\mc,h}^\coins}{\coinsorder}$. We fix a total order on states. Then  we fix \emph{state/step coins} $\coins=\{ c_{s,i} \mid s \in S, i < h \}$ and the weights as in Example~\ref{ex:pathspredicate}. Finally, let $\coinsorder$ be an order on $C$ such that $i < j$ implies $c_{s,i} \coinsorder c_{s,j}$. Then:
\begin{equation}
  \ct{\mc}{h} \bisim  \bddw{\varphi_{\mc,h}^\coins}{\coinsorder}{\coinsweight}.
\end{equation}
In the spirit of Idea~1, we thus aim to construct $ \bddw{\varphi_{\mc,h}^\coins}{\coinsorder}{\coinsweight}$, a representation as outlined in Idea~2, efficiently.
Indeed, the BDD (as MC) in Fig.~\ref{fig:ctbdd} is bisimilar to the MC in Fig.~\ref{fig:compressedct}. 

\begin{mdframed}
  \textbf{Idea 3:} Represent a bisimilar version of the computation tree using a BDD.
\end{mdframed}    


\subsection{The Algorithmic Benefits of BDD Construction}
\label{sec:causalitys}
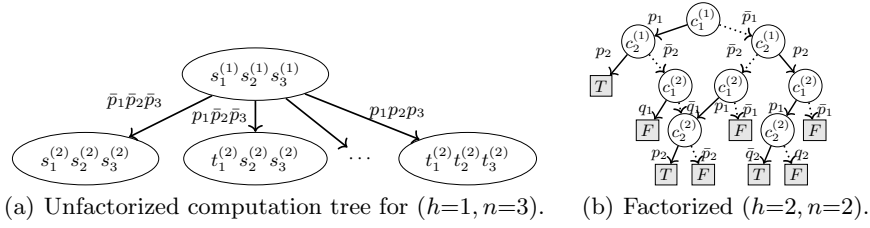
\begin{figure}[t]
  \centering
    \subfigure[Unfactorized computation tree for $(h{=}1,n{=}3)$.]{
    \scalebox{0.8}{
      \begin{tikzpicture}
    \def\lvl{40pt}
    \node[draw, ellipse] (s000) at (0bp,0bp) {$s_1^{(1)}s_2^{(1)}s_3^{(1)}$};
    \node[draw, ellipse] (p000) at (-80bp,-\lvl) {$s_1^{(2)}s_2^{(2)}s_3^{(2)}$};
    \node[draw, ellipse] (p100) at (0bp,-\lvl) {$t_1^{(2)}s_2^{(2)}s_3^{(2)}$};
    \node (dots) at (50bp, -\lvl) {$\cdots$};
    \node[draw, ellipse] (p111) at (100bp,-\lvl) {$t_1^{(2)}t_2^{(2)}t_3^{(2)}$};

    \begin{scope}[on background layer]
       \draw [thick, ->] (s000) -- (p000) node[midway,above left] {$\bar{p}_1\bar{p}_2\bar{p}_3$};
       \draw [thick, ->] (s000) -- (p100) node[midway, left] {$p_1\bar{p}_2\bar{p}_3$};
       \draw [thick, ->] (s000) -- (dots);
       \draw [thick, ->] (s000) -- node[right,yshift=2pt] {$p_1{p}_2{p}_3$} (p111);

    \end{scope}
  \end{tikzpicture}
  \label{fig:unfactorized}
}
}
   ~ 
  \subfigure[Factorized $(h{=}2,n{=}2)$.]{
    \scalebox{0.7}{
      \begin{tikzpicture}
    \def\lvl{24pt}
    \node (s00) at (0bp,0bp) [bddnode] {$c_1^{(1)}$};

    \node (s10a) at ($(s00) + (-35bp, -0.5*\lvl)$) [bddnode] {$c_2^{(1)}$};
    \node (s10b) at ($(s00) + (35bp, -0.5*\lvl)$) [bddnode] {$c_2^{(1)}$};

    \node (t0) at ($(s10a) + (-20bp, -\lvl)$) [bddterminal] {$T$};
    \node (s01a) at ($(s10a) + (20bp, -\lvl)$) [bddnode] {$c_1^{(2)}$};
    \node (s01b) at ($(s10b) + (20bp, -\lvl)$) [bddnode] {$c_1^{(2)}$};
    \node (s01c) at ($(s10b) + (-20bp, -\lvl)$) [bddnode] {$c_1^{(2)}$};

    \node (s11a) at ($(s01a) + (5bp, -\lvl)$) [bddnode] {$c_2^{(2)}$};
    \node (f0) at ($(s01a) + (-15bp, -\lvl)$) [bddterminal] {$F$};
    \node (s11b) at ($(s01b) + (-15bp, -\lvl)$) [bddnode] {$c_2^{(2)}$};
    \node (f1) at ($(s01b) + (5bp, -\lvl)$) [bddterminal] {$F$};
    \node (f2) at ($(s01c) + (5bp, -\lvl)$) [bddterminal] {$F$};

    \node (t11a) at ($(s11a) + (-10bp, -\lvl)$) [bddterminal] {$T$};
    \node (f11a) at ($(s11a) + (10bp, -\lvl)$) [bddterminal] {$F$};
    \node (t11b) at ($(s11b) + (-10bp, -\lvl)$) [bddterminal] {$T$};
    \node (f11b) at ($(s11b) + (10bp, -\lvl)$) [bddterminal] {$F$};
\node[right=0.8cm of f11b] {};

    \begin{scope}[on background layer]
       \draw [thick, ->] (s00) -- (s10a) node[midway,above left] { $p_1$};
       \draw [thick, lowedge, ->] (s00) -- (s10b) node[midway,above right] {$\bar{p}_1$};
       \draw [thick, ->] (s10a) -- (t0) node[midway,above left] {$p_2$};
       \draw [thick, lowedge, ->] (s10a) -- (s01a) node[midway,above right] {$\bar{p}_2$};
       \draw [thick, ->] (s01a) -- (f0) node[midway, left] {${q}_1$};
       \draw [thick, lowedge, ->] (s01a) -- (s11a) node[midway, right] {$\bar{q}_1$};
       \draw [thick, ->] (s11a) -- (t11a) node[midway, left] {${p}_2$};
       \draw [thick, lowedge, ->] (s11a) -- (f11a) node[midway, right] {$\bar{p}_2$};

       \draw [thick, ->] (s10b) -- (s01b) node[midway, above right] {${p}_2$};
       \draw [thick, lowedge, ->] (s10b) -- (s01c) node[midway, above left] {$\bar{p}_2$};
       \draw [thick, ->] (s01b) -- (s11b) node[midway,  left] {${p}_1$};
       \draw [thick, lowedge, ->] (s01b) -- (f1) node[midway, right] {$\bar{p}_1$};
       \draw [thick,  ->] (s11b) -- (t11b) node[midway, left] {$\bar{q}_2$};
       \draw [thick, lowedge, ->] (s11b) -- (f11b) node[midway, right] {${q}_2$};

       \draw [thick, ->] (s01c) -- (s11a) node[midway, right] {${p}_1$};
       \draw [thick, lowedge, ->] (s01c) -- (f2) node[midway, right] {$\bar{p}_1$};
    \end{scope}

  \end{tikzpicture}
}
      \label{fig:twochaincollapsed}
    }

    \caption{Two computation trees for the motivating example in Section~\ref{sec:motivating_example}.}
    \label{fig:bdds}
\end{figure}

Thus far we have described how to construct a binarized MC 
bisimilar to the CT. Here, we argue that this construction has
algorithmic benefits by filling in two details. First, the 
binarized representation is an important ingredient for compact BDDs. Second, we
show how to choose a variable ordering that ensures that the  BDDs
grow linearly in the horizon. In sum,
\begin{mdframed}
  \textbf{Idea 4:} $\WMC$ encodings of binarized Markov Chains may increase
compression of computation trees.
\end{mdframed}

To see the benefits of binarized transitions, we return to the 
factory example in Section~\ref{sec:motivating_example}.
Figure~\ref{fig:unfactorized} gives a bisimilar computation tree for the 3-factory
$h=1$ example. However, in this tree, the states are \emph{unfactorized}: each
node in the tree is a joint configuration of factories. This tree has 8
transitions (one for each possible joint state transition) with 8 distinct
probabilities. On the other hand, the bisimilar computation tree in
Figure~\ref{fig:motiv} has binarized transitions: each node corresponds to a
single factory's state at a particular time-step, and each transition describes
an update to only a single factory.
This binarization enables the exploitation
of new structure: in this case, the independence of the factories leads to
smaller BDDs, that is otherwise lost when considering only joint configurations
of factories.

Recall that the size of the ADD representation of the transition function is
bounded from below by the number of distinct probabilities in the underlying MC: in this case, this
is visualized by the number of distinct outgoing edge probabilities from all nodes in the
unfactorized computation tree. Thus, a good binarization can have a
drastically positive effect on performance. For the running
example, rather than $2^{n}$ different transition probabilities (with $n$ factories), the system
now has only $4\cdot n$ distinct transition probabilities! 

\myparagraph{Causal orderings.}
Next, we explore some of the \emph{engineering choices} \rubicon{} makes to exploit the sequential structure in a MC when
constructing the BDD for a $\WMC$ query. First, note that the transition matrix
$P(s, s')$ implicitly encodes a distribution over state transition functions, $S \to S$.
To encode $P$ as a BDD, we must encode each transition as a logical
variable, similar to the situation in \secref{logical}. In the case of binary
transitions this is again easy. In the case of non-binary transitions, we 
again introduce additional logical variables \cite{fierens2015inference,sang2005performing,HoltzenOOPSLA20,chavira2008probabilistic}.
This logical function has the following form:
\begin{equation}
  f_P \colon \bool^{\coins} \to (S \to S).
\end{equation}

Whereas the computation tree follows a fixed (temporal) order of states, BDDs
can represent the same function (and the same weighted model count) using an
arbitrary order. Note that the BDD's size and structure drastically depends both on the
construction of the propositional formula \emph{and} the order of the variables
in that encoding. We can bound the size of the BDD by enforcing a variable order based on  the temporal structure of the original
MC. Specifically, given $h$ coin collections $\vec{C} = C
\times \ldots \times C$, one can generate a function $f$ describing the $h$-length paths via repeated
applications of $f_P$:
 \begin{equation}\label{eq:causal_encoding}
   \begin{split}
     &f \colon \bool^{\vec{C}} \to \hPaths\quad f(\coins_1, \ldots, \coins_h) = \bigg ( f_P(\coins_h)\circ \ldots \circ f_P(\coins_1)\bigg )(\iota)
   \end{split}
 \end{equation}
 Let $\psi$ denote an indicator for the reachability property as a function over paths, $\psi\colon \hPaths \to \bool$ with $\psi(\pi)= [ \pi \in \seteventuallyb{h}{T}]$. 
 We call predicates formed by composition with $f_P$, i.e.,
 $\varphi = \psi \circ f_P$, \emph{causal encodings} and
 orderings on $c_{i, t} \in \vec{C}$ that are
 lexicographically sorted in time,
 $t_1 < t_2 \implies c_{i, t_1} < c_{j, t_2}$, \emph{causal orderings}. Importantly,
 causally ordered / encoded BDDs grow linearly in horizon
 $h$~\cite[Corollary 1]{DBLP:conf/cav/VazquezChanlatte20}.  More
 precisely, let $\varphi_{\mc,h}^{\vec{\coins}}$ be causally encoded
 where $|\vec{C}| = h\cdot m$. The causally ordered BDD for
 $\varphi_{\mc,h}^{\vec{\coins}}$ has at most
 $h\cdot |S\times S_\psi| \cdot m\cdot2^{m}$ nodes, where $|S_\psi| = 2$ for reachability properties.\footnote{Generally, it is the smallest number of states required for a DFA to recognize $\psi$.}
However, while the worst-case growth is linear in the
 horizon, constructing that BDD may induce a super-linear cost in the
 size, e.g., function composition using BDDs is super-linear!

Figure~\ref{fig:twochaincollapsed} shows the motivating factory example with 2
factories and $h=2$. The variables are causally ordered: the
factories in time step 1 occur before the factories in time step 2. For $n$ 
 factories, a fixed number $f(n)$ of nodes are added to the BDD upon each
iteration, guaranteeing growth on the order $\mathcal{O}(f(n)\cdot h)$. Note the factorization that occurs: the
BDD has node sharing (node $c_2^{(2)}$ is reused) that yields additional computational benefits.

\emph{Summary and remaining steps.}
The operational view highlights that we want to compute a transformation of the original input MC $\mc$. 
The logical view presents an approach to do so efficiently: By computing a BDD that stores a predicate describing all paths that reach the target, and interpreting and evaluating the (graph of the) BDD as an MC.
In the following section, we discuss the two steps that we follow to create the BDD: 
\begin{enumerate*}[label=(\roman*)]
	\item From $\prog$ generate $\prog'$ such that $\ct{\semantics{\prog}}{h} \bisim \semantics{\prog'}$.
	\item From $\prog'$ generate $\mc$ such that $\mc = \semantics{\prog'}$.
\end{enumerate*}

%
%
%
 

%
%
%
\section{\rubicon}
\label{sec:translation}

We present \rubicon{} which follows the two steps outlined above.
For
exposition, we first describe a translation of \emph{monolithic} \prism{} programs to \dice{} programs
and then extend this translation to admit
modular programs. Technical steps and extensions are deferred to Appendix~\ref{sec:translation:extensions}.


\mysubsubsection{\dice{} Preliminaries}
We give a brief description of \dice{}, a probabilistic programming
language (PPL) introduced in \cite{HoltzenOOPSLA20}.
A PPL is a programming language augmented with a
primitive notion of random choice: for instance, in \dice{}, a Bernoulli random
variable is introduced by the syntax \texttt{flip 0.5}. The syntax of \dice{} is
similar to the programming language \texttt{OCaml}: local variables are
introduced by the syntax \texttt{let x = e$_1$ in e$_2$}, where \texttt{e}$_1$
and \texttt{e}$_2$ are \emph{expressions}, i.e.,
sub-programs. \dice{} supports procedures, bounded integers, bounded loops, and standard
control flow via \texttt{if}-statements.

One goal of a PPL is to perform \emph{probabilistic inference}: compute the
probability that the program returns a particular value. Inference
on the tiny \dice{} program \texttt{let x = flip 0.1 in x} would yield that  \texttt{true} is returned with probability $0.1$. 
The \dice{} compiler performs
probabilistic inference via weighted model counting and BDD compilation. In
doing so, it accomplishes the \emph{non-trivial} tasks of:
\begin{enumerate*}[label=(\roman*)]
  \item choosing a logical encoding for probabilistic programs
  \item establishing good variable orderings
  \item efficiently manipulating and constructing BDDs
  \item performing WMC
\end{enumerate*}. For details, we refer
the reader to \cite{HoltzenOOPSLA20}. 

\rubicon{} uses \dice{} to effectively construct a BDD and perform WMC on a
\dice{} program that reflects a description of some computation tree.
This implementation exploits the structure that was described in
\secref{causalitys}: in particular, the BDD generated in
Figure~\ref{fig:twochaincollapsed} is exactly the BDD that will be generated by
\dice{} from the output of \rubicon{}. 
The variable ordering used by \dice{} is given by the order in which program
variables are introduced, and \rubicon{}'s translation was designed with this
variable ordering in mind.

\mysubsubsection{Transpiling \prism{} to \dice{}}
We present the core translation routine implemented in  \rubicon{}. We note that the  ultimate
performance of \rubicon{} is heavily dependent on the quality of this 
translation.  We evaluate the performance in the next section.

The \prism{} specification language consists of one or more reactive \emph{modules} (or partially synchronized state machines) that may
interact with each other. Our example in 
Fig.~\ref{fig:motivprism} illustrates fully synchronized state
machines. While \prism{} programs containing multiple modules can be
flattened into a single monolithic program, this yields an exponential blow-up:
If one flattens the $n$ modules in Fig.~\ref{fig:motivprism} to a single module,
the resulting program has $2^n$ updates per command. This motivates our direct
translation of PRISM programs containing multiple modules.

\lstset{language=Prism}   
\newsavebox{\programbox}
\begin{lrbox}{\programbox}
\begin{lstlisting}[numbers=none]
module main
  x : [0..1] init 0;
  y : [0..2] init 1;
  [] x=0 & y<2 -> 0.5:x'=1 + 0.5:y'=y+1;       
  [] y=2 -> 1:y'=y-1; 
  [] x=1 & y!=2 -> 1:x'=y & y'=x;
endmodule
property: P=? [F<=2 (x=0 & y=2)]
\end{lstlisting}
\end{lrbox}
\newsavebox{\prismpd}
\begin{lrbox}{\prismpd}
\begin{lstlisting}[numbers=none]
module main
x : [0..2] init 0;
[] true -> x/(x+1):x'=x-1 +            
           1-(x/(x+1)):x'=x+1; 
endmodule
\end{lstlisting}
\end{lrbox}
\newsavebox{\prismog}
\begin{lrbox}{\prismog}
\begin{lstlisting}[numbers=none]
module main
x : [0..2] init 1;
y : [0..2] init 1;
[] x>1 -> 1:x'=y&y'=x;
[] y<2 -> 1:x'=min(x+1,2); 
endmodule
\end{lstlisting}
\end{lrbox}
\lstset{language=dice}   
\newsavebox{\dicestructfull}
\begin{lrbox}{\dicestructfull}
\begin{lstlisting}
s   =  init()
if !valid(s) then ERR else
s, T =  step(s), hit(s)
if !valid(s) then ERR else
s, T =  if !T then step(s), hit(s)
       else       s, T
if !valid(s) then ERR else
s, T =  if !T then step(s), hit(s)
       else       res, T
T
\end{lstlisting}
\end{lrbox}
\newsavebox{\dicestruct}
\begin{lrbox}{\dicestruct}
\begin{lstlisting}
let s =  init() in // init state
let T =  hit(s) in   // init target
let (s, T) =  if !T 
   then let s' =  step(s) in (s', hit(s')) 
   else (s, T) in
let (s, T) =  if !T then 
   then let s' =  step(s) in (s', hit(s')) 
   else (s, T) in
T
\end{lstlisting}
\end{lrbox}

\newsavebox{\dicevanillafull}
\begin{lrbox}{\dicevanillafull}
\begin{lstlisting}
fun init() { (0,1) }
fun valid(s:(x,y)) 
{ x <= 1 && y <= 2 }
fun hit((x,y)) 
{ x==0 && y==2 }
fun step((x,y) ) {
  if x==0 && y<2 then
   if flip 0.5 then (1,y) else (x,y+1)
  else if y == 2 then (x,y-1)
  else if x==1 && y != 2 then (y,x)
  else (x,y) } 
\end{lstlisting}
\end{lrbox}
\newsavebox{\dicevanilla}
\begin{lrbox}{\dicevanilla}
\begin{lstlisting}
fun init() { (0,1) }
fun hit((x,y)) { x ==  0  &&  y ==  2 }
fun step((x,y)) {
  if x==0 &&  y<2 then
     if flip 0.5   then (1,y) else (x,y+1)
  else if y==2 then (x,y-1)
  else if x==1 &&  y!=1 then (y,x)
  else (x,y) 
} 
\end{lstlisting}
\end{lrbox}

\begin{figure}[t]
  \centering
  
  \subfigure[\prism{} program with reachability query]{ \raisebox{20pt}{\usebox{\programbox}}
    \label{fig:prismprogram}
  }
  \subfigure[Underlying MC]{ \begin{tikzpicture}[ every node/.style={scale=0.9, font=\scriptsize}, st/.style={circle, inner sep=0.5pt, minimum size=13pt},baseline=(s10)]

	\node[st] (s00) {$\langle 0, 0\rangle$};
	\node[st, right=1.1cm of s00,initial,initial text=,initial where=above,initial distance=3mm] (s01) {$\langle 0, 1\rangle$};
	\node[st, right=1.1cm of s01] (s02) {$\langle 0, 2\rangle$};
	\node[st, below=0.5cm of s00] (s10) {$\langle 1, 0\rangle$};
	\node[st, right=1.1cm of s10] (s11) {$\langle 1, 1\rangle$};
	\node[st, right=1.1cm of s11] (s12) {$\langle 1, 2\rangle$};

	
	\draw[->] (s00) edge node[above] {$\nicefrac{1}{2}$} (s01);
	\draw[->] (s00) edge node[left] {$\nicefrac{1}{2}$} (s10);
	\draw[->] (s01) edge[bend left] node[below] {$\nicefrac{1}{2}$} (s02);
	\draw[->] (s01) edge node[left] {$\nicefrac{1}{2}$} (s11);
		
	\draw[->] (s02) edge[bend left] node[above] {$1$} (s01);
	\draw[->] (s12) edge node[above] {$1$} (s11);

	\draw[->] (s10) edge node[right] {$1$} (s01);


	\draw[->] (s11) edge[loop left] node[left] {$1$} (s11);

\end{tikzpicture}%
    \label{fig:underlyingmodel}
  }
  \subfigure[Main \dice{} program for $h{=}2$]{ \usebox{\dicestruct}
    \label{fig:dicestruct}
  }
  \quad
  \subfigure[\dice{} auxiliary functions]{ \usebox{\dicevanilla}
    \label{fig:vanilladice}
  }
  \caption{From \prism{} to \dice{} using \rubicon.}
\end{figure}

\myparagraph{Monolithic Prism programs.}
We explain most ideas on \prism{} programs that consist of a single
``monolithic'' module before we address the modular translation at the end of
the subsection. A module has a set of bounded variables, and the valuations of these
variables span the state space of the underlying MC. Its transitions are
described by guarded \emph{commands} of the form:
\[
[\texttt{act}]\ \ \mbox{\texttt{guard}} 
\ \ \rightarrow \ \ 
p_1 : \mbox{\texttt{update}}_1 + \ldots \ldots + p_n : \mbox{\texttt{update}}_n 
\]
The \emph{action} name \texttt{act} is only relevant in the modular case and can be ignored for now.
The \emph{guard} is a Boolean expression over the module's  variables. 
If the guard evaluates to $\true$ for some state (a valuation), then the module evolves into one of the $n$ successor states by updating its variables.
An \emph{update} is chosen according to the probability distribution given by the expressions $p_1, \hdots, p_n$.
In every state enabling the guard, the evaluation of $p_1,\hdots,p_n$ must sum up to one.
A set of guards \emph{overlap} if they all evaluate to $\true$ on a given state.
The semantics of overlapping guards in the monolithic setting is to first uniformly select an active guard
and then apply the corresponding stochastic transition. Finally, a self-loop is implicitly added to states without an enabled guard. 
\begin{example}
  We present our translation primarily through example.
  In Fig.~\ref{fig:prismprogram}, we give a \prism{} program for a
  MC. The program contains two variables $x$ and $y$, where $x$ is
  either zero or one, and $y$ between zero and two. There are thus 6
  different states. We denote states as tuples with the $x$- and
  $y$-value. We depict the MC in
  Fig.~\ref{fig:underlyingmodel}.  From state $\langle 0, 0 \rangle$,
  (only) the first guard is enabled and thus there are two
  transitions, each with probability a half: one in which $x$ becomes
  one and one in which $y$ is increased by one.
  Finally, there is no guard enabled in state $\langle 1, 1 \rangle$, resulting
  in an implicit self-loop.
\end{example}
\myparagraph{Translation.}
All \dice{} programs consist of two parts: a \emph{main} routine, which is run
by default when the program starts, and \emph{function declarations} that
declare auxiliary functions. We first define the auxiliary functions. For
simplicity let us temporarily assume that no guards overlap and that
probabilities are constants, i.e., not state-dependent. 

The main idea in the
translation is to construct a \dice{} function \texttt{step} that, given
the current state, outputs the next state. Because a monolithic \prism{} program
is almost a sequential program, in its most basic version, the \texttt{step}
function is straightforward to construct using built-in \dice{} language
primitives: we simply build a large if-else block corresponding to each command.
This block iteratively considers each command's guard until it finds one that is
satisfied. To perform the corresponding update we flip a coin -- based on the
probabilities corresponding to the updates -- to determine which update to
perform. If no command is enabled, we return the same state in accordance with
the implicit self-loop. Fig.~\ref{fig:vanilladice} shows the program blocks for
the \prism{} program from Fig.~\ref{fig:prismprogram} with target state
$\semantics{x=0,y=2}$. There are two other important auxiliary functions. The
\texttt{init} function simply returns the initial state by translating the
initialization statements from \prism{}, and the \texttt{hit} function checks
whether the current state is a target state that is obtained from the property.

Now we outline the main routine, given for this example in
Figure~\ref{fig:dicestruct}. This function first initializes
the state. Then, it calls \texttt{step} $2$ times, checking on each iteration using
\texttt{hit} if the target state is reached. Finally, we return whether we have
been in a target state. The probability to return true corresponds to the
reachability probability on the underlying MC specified by the \prism{} program.

\newsavebox{\diceog}
\begin{lrbox}{\diceog}
\begin{lstlisting}
fun step((x,y)) {
  let aEn = (x>1)                      in
  let bEn = (y<2)                      in
  let act =  selectFrom(aEn, bEn)     in 
  if act==1 then (y,x)
  else if act==2 then (min(x+1,2),y)
  else (x,y)} ...
\end{lstlisting}
\end{lrbox}
\newsavebox{\dicepd}
\begin{lrbox}{\dicepd}
\begin{lstlisting}
fun step(x) {
  if true then
    if x==0 then
      if flip 0 then x-1 else x+1
    else if x==1 then 
      if flip 0.5 then x-1 else x+1
    else
      if flip 0.33 then x-1 else x+1
} ...
\end{lstlisting}
\end{lrbox}

\begin{figure}[t]
\centering

\subfigure[]{
\usebox{\prismog}
\label{fig:prismog}
}
\qquad
\subfigure[]{
\usebox{\diceog}
\label{fig:diceog}
}

\vspace{-1mm}
\caption{\prism{} program with overlapping guards and its translation (conceptually).}	
\end{figure}

\myparagraph{Overlapping guards.}
\prism{} allows multiple commands to be enabled
in the same state, with semantics to uniformly at random
choose one of the enabled commands to evaluate. 
\dice{} has no primitive notion of this construct.\footnote{One cannot simply condition on selecting an
enabled guard as this redistributes probability mass over all paths and not only
over paths with the same prefix.}
We illustrate the translation in Fig.~\ref{fig:prismog} and Fig.~\ref{fig:diceog}.
It determines which guards \texttt{aEn}, \texttt{bEn}, \texttt{cEn} are enabled. Then, we randomly select one of the commands which are enabled, i.e., we uniformly at random select a true bit from a given tuple of bits. We store the index of that bit and use it to execute the corresponding command.

\lstset{language=Prism}   
\newsavebox{\modularprism}
\begin{lrbox}{\modularprism}
\begin{lstlisting}[numbers=none]
module m1
x : [0..1]  init 0;
[a] x=1 ->  1:x'=1-y;
[b] x=0 ->  1:x'=0; 
endmodule
module m2
y : [0..1]  init 0;
[b] y=1 ->  0.5:y'=0 + 0.5:y'=1; 
[c] true ->  1:x'=1-x;
endmodule
\end{lstlisting}
\end{lrbox}
\lstset{language=Dice}   
\newsavebox{\modulardice}
\begin{lrbox}{\modulardice}
\begin{lstlisting}[numbers=none]
fun step((x,y)) {
  let aEn = (x==1) in 
  let bEn = (x=0 && y=1) in 
  let cEn = true in
  let act = selectFrom(aEn, bEn, cEn) in
  if act==1 then (1-y, y)
  else if act==2 then (0,   flip 0.5)
  else if act==3 then (1-x, y)
  else (x,   y)
}
\end{lstlisting}
\end{lrbox}

\begin{figure}[t]

    \centering
    \subfigure[]{
    \usebox{\modularprism}
    }
    \subfigure[]{
    \usebox{\modulardice}
    }
\vspace{-1mm}
    \caption{Modular \prism{} and resulting \dice{} step function.}
    \label{fig:modulartranslation}
\end{figure}

\myparagraph{Modular Prism Programs.}
For modular \prism{} programs, the \emph{action names} at the front of \prism{} commands  are important.  In each module, there is a set of action names available. An action is \emph{enabled} if each module that contains this action name has (at least) one command with this action whose guard is satisfied.
Commands with an empty action are assumed to have a globally unique action name, so in that case the action is enabled iff the guard is enabled.
Intuitively, once an action is selected, we randomly select a command per module in all modules containing this action name.  
Our approach resembles that for overlapping guards described above.
See  Fig.~\ref{fig:modulartranslation} for an intuitive example.
To automate this, the updates require more care, see Appendix~\ref{sec:translation:extensions} for details.

\myparagraph{Implementation.}
\rubicon{} is implemented on top of \storm's Python API and translates \prism{} to \dice{} fully automatically. 
\rubicon{} supports all MCs in the \prism{} benchmark suite 
  and a large set of benchmarks from the \prism{} website and the QVBS~\cite{DBLP:conf/tacas/HartmannsKPQR19}, with the note that we require a single initial state and ignore reward declarations. Furthermore, we currently do not support the hide/restrict
process-algebraic compositions and some integer operations.


\section{Empirical Comparisons}
\label{sec:experiments}

We compare and contrast the performance of
\storm{} against \rubicon{} 
to empirically demonstrate the following strengths and
weaknesses:\footnote{All
  experiments were conducted with \storm{} version 1.6.0 on the same server with
  512GB of RAM, using a single thread of
  execution. Time was reported using the built-in Unix \texttt{time} utility;
  the total wall-clock time is reported.}
\begin{description}
\item[Explicit Model Checking (\storm{})] represents the MC explicitly in a sparse matrix  format. The approach suffers from the state space explosion, but has been engineered to scale to models with many states. Besides the state space, the sparseness of the transition matrix is essential for performance.
\item[Symbolic Model Checking (\storm{})] represents the transition matrix and the reachability probability as an
  ADD. This method is strongest when the transition matrix and state vector have structure
  that enables a small ADD representation, like symmetry and sparsity.
\item[\rubicon{}] represents the set of paths through the MC as a
  (logical) BDD. This method excels when the state space
  has structure that enables a compact BDD representation, such as conditional
  independence, and hence scales well on examples with many (asymmetric) parallel
  processes or queries that admit a compact representation. 
\end{description}
The sources, benchmarks and binaries are archived.\footnote{\url{doi.org/10.5281/zenodo.4726264} and \url{github.com/sjunges/rubicon}}

There is no
clear-cut model checking technique that is superior to others (see QCOMP~\cite{qcomp}). We
demonstrate that, while \rubicon{} is not competitive on some
commonly used benchmarks~\cite{DBLP:conf/qest/KwiatkowskaNP12}, it improves a modern model checking
portfolio approach on a significant set of benchmarks.
Below we provide several natural
models on which \rubicon{} is superior to one or both competing methods.
We also evaluated \rubicon{} on standard benchmarks, highlighting that \rubicon{} is applicable to models
from the literature. We see that \rubicon{} is effective on \textsc{Herman}
(elaborated below), has mixed results on \textsc{BRP} (see Appendix~\ref{app:exp}) 
 and is currently not competitive on some other standard benchmarks (NAND, EGL, LeaderSync).
While not exhaustive, our selected benchmarks highlight specific strengths and weaknesses of \rubicon.
Finally, a particular benefit of \rubicon{} is fast sampling of
parametric chains, which we demonstrate on 
\textsc{Herman} and our factory example.



\begin{figure}[t]
  \centering
\subfigure[Weather Factory]{
   \begin{tikzpicture}
   \begin{axis}[
    xlabel near ticks,
		height=2.9cm,
   width=3.5cm,
   ylabel={Time (s)},
   xlabel style = {font=\scriptsize},
   ylabel style = {font=\scriptsize},
   ylabel near ticks,
		grid=major,
   xlabel={\# Factories},
   ticklabel style = {font=\tiny},
   legend style={at={(0em,10em)},anchor=north},
   ]
   \addplot[mark=square, thick] table [x index={0}, y index={1}] {\weather};\label{plt:rubicon1}
   \addplot[mark=diamond, thick, red] table [x index={0}, y index={2}] {\weather}; \label{plt:stormsym1}
   \addplot[mark=star, thick, green] table [x index={0}, y index={3}] {\weather}; \label{plt:stormexp1}

 \end{axis}
\end{tikzpicture}
 \label{fig:expweather}
}~
\subfigure[{Weather~Factory~2}]{
   \begin{tikzpicture}
   \begin{axis}[
    xlabel near ticks,
		height=2.9cm,
   width=3.7cm,
   xlabel style = {font=\scriptsize},
		grid=major,
   xlabel={\# Factories},
   ticklabel style = {font=\tiny},
   legend style={at={(0em,10em)},anchor=north},
   ]
   \addplot[mark=square, thick] table [x index={0}, y index={1}] {\weathertwo};
   \addplot[mark=diamond, thick, red] table [x index={0}, y index={3}] {\weathertwo}; 
   \addplot[mark=star, thick, green] table [x index={0}, y index={2}] {\weathertwo}; 

 \end{axis}
\end{tikzpicture}
 \label{fig:expweather2}
}~
  \subfigure[Herman-13]{
 \centering
     \begin{tikzpicture}
   \begin{axis}[
    xlabel near ticks,
		height=2.9cm,
   width=3.5cm,
   ticklabel style = {font=\tiny},
   xlabel style = {font=\scriptsize},
		grid=major,
   xlabel={Horizon $(h)$},
   ]
   \addplot[mark=square, thick] table [x index={0}, y index={1}] {\hermanthirteen};
   \addplot[mark=diamond, thick, red] table [x index={0}, y index={2}] {\hermanthirteen}; 
   \addplot[mark=star, green, thick] table [x index={0}, y index={4}] {\hermanthirteen};

 \end{axis}
\end{tikzpicture}
\label{fig:expherman13}
}~
  \subfigure[Herman-13 (R)]{
 \centering
     \begin{tikzpicture}
   \begin{axis}[
    xlabel near ticks,
		height=2.9cm,
   width=3.5cm,
   ticklabel style = {font=\tiny},
   xlabel style = {font=\scriptsize},
		grid=major,
   xlabel={Horizon $(h)$},
   ]
   \addplot[mark=square, thick] table [x index={0}, y index={1}] {\hermanthirteen};
   \addplot[mark=diamond, blue, red, thick] table [x index={0}, y index={3}] {\hermanthirteen}; 
   \addplot[mark=star, green, thick] table [x index={0}, y index={4}] {\hermanthirteen};

 \end{axis}
\end{tikzpicture}
\label{fig:expherman13random}
}
\\

\vspace{-2mm}
\subfigure[Herman-17]{
\begin{tikzpicture}
   \begin{axis}[
		height=2.9cm,
   width=3.5cm,
    xlabel near ticks,
   ticklabel style = {font=\tiny},
   ylabel style = {font=\scriptsize},
   ylabel={Time (s)},
   xlabel style = {font=\scriptsize},
      ylabel near ticks,
		grid=major,
   xlabel={Horizon $(h)$},
   ]
   \addplot[mark=square, thick] table [x index={0}, y index={1}] {\hermanseventeen};
   \addplot[mark=diamond, thick, red] table [x index={0}, y index={2}] {\hermanseventeen}; 
   \addplot[mark=star, green, thick] table [x index={0}, y index={4}] {\hermanseventeen}; 

 \end{axis}
\end{tikzpicture}
\label{fig:expherman17}
}~
\subfigure[Herman-17 (R)]{
\begin{tikzpicture}
   \begin{axis}[
		height=2.9cm,
   width=3.5cm,
       xlabel near ticks,
   ticklabel style = {font=\tiny},
   xlabel style = {font=\scriptsize},
      ylabel near ticks,
		grid=major,
   xlabel={Horizon $(h)$},
   ]
   \addplot[mark=square, thick] table [x index={0}, y index={1}] {\hermanseventeen};
   \addplot[mark=diamond, thick, red] table [x index={0}, y index={3}] {\hermanseventeen};
   \addplot[mark=star, green, thick] table [x index={0}, y index={4}] {\hermanseventeen}; 

 \end{axis}
\end{tikzpicture}
\label{fig:expherman17random}
}
~
\subfigure[Herman-19 (R)]{
\begin{tikzpicture}
   \begin{axis}[
		height=2.9cm,
   width=3.3cm,
       xlabel near ticks,
   ticklabel style = {font=\tiny},
   xlabel style = {font=\scriptsize},
		grid=major,
   xlabel={Horizon $(h)$},
   ]
   \addplot[mark=square, thick] table [x index={0}, y index={1}] {\hermannineteen};
   \addplot[mark=star, green, thick] table [x index={0}, y index={2}] {\hermannineteen}; 

 \end{axis}
\end{tikzpicture}
\label{fig:expherman19}
}
\subfigure[Queues]{
\begin{tikzpicture}
   \begin{axis}[
		height=3cm,
   width=3.5cm,
       xlabel near ticks,
   ticklabel style = {font=\tiny},
   xlabel style = {font=\scriptsize},
		grid=major,
   xlabel={Horizon $(h)$},
   ]
   \addplot[mark=square, thick] table [x index={0}, y index={1}] {\queues};
   \addplot[mark=diamond, thick, red] table [x index={0}, y index={2}] {\queues};
   \addplot[mark=star, green, thick] table [x index={0}, y index={3}] {\queues}; 

 \end{axis}
\end{tikzpicture}
\label{fig:queues}
}
\vspace{-2mm}
\caption{Scaling plots comparing \rubicon{} (\ref{plt:rubicon1}), \storm{}'s
  symbolic engine (\ref{plt:stormsym1}), and \storm{}'s explicit engine (\ref{plt:stormexp1}).
An ``(R)'' in the caption denotes random parameters.
}
\label{fig:scaling}
\end{figure}
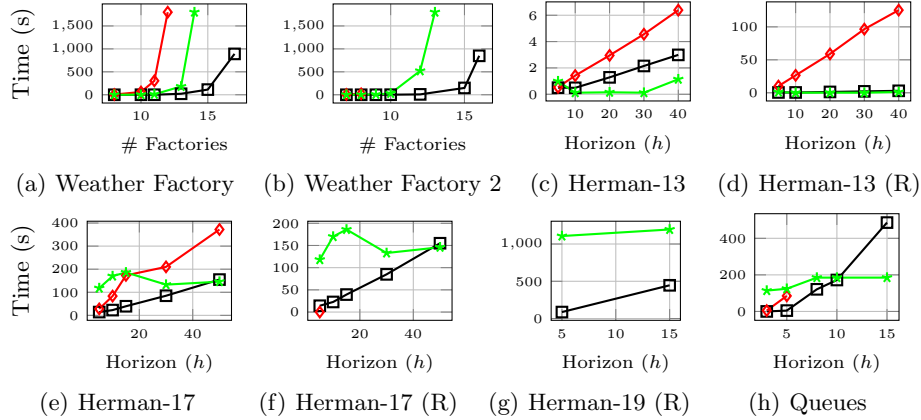

\mysubsubsection{Scaling Experiments}
\label{sec:expscaling}
In this section, we describe several scaling experiments
(Figure~\ref{fig:scaling}), each designed to highlight a specific
strength or weakness.

\emph{Weather Factories.}
First, Figure~\ref{fig:expweather} describes a generalization of the motivating
example from \secref{motivating_example}. In this model, the
probability that each factory is on strike is dependent on a common random
event: whether or not it is raining. The rain on each day is
dependent on the previous day's weather. We plot runtime for an increasing number of factories for $h{=}10$. 
Both \storm{} engines  eventually fail due to the state explosion and the number of distinct probabilities in the MC.
\rubicon{} is orders
of magnitude faster in comparison, highlighting that it does not depend on complete independence among the factories.
Figure~\ref{fig:expweather2} shows a more challenging instance where the weather includes \emph{wind} which, each day,
affects whether or not the sun will shine, which in turn affects strike
probability.

\emph{Herman}. Herman is based on a distributed
protocol~\cite{DBLP:journals/ipl/Herman90} that has been
well-studied~\cite{DBLP:journals/fac/KwiatkowskaNP12,DBLP:conf/srds/AflakiVBKS17}
and which is one of the standard benchmarks in probabilistic model checking.
Rather than computing the expected steps to `stabilization', we consider the step-bounded probability of stabilization. 
Usually, all participants in the protocol flip a coin with the same bias.
The model is then highly symmetric, and hence is amenable to symbolic representation with ADDs.
Figures~\ref{fig:expherman13} and \ref{fig:expherman17} show how the methods
scale on Herman examples with 13 and 17 parallel processes. We observe that the explicit approach scales very efficiently in the number of
iterations but has a much higher up-front model-construction cost, and hence can
be slower for fewer iterations. 

To study what happens when the coin biases vary over the protocol participants,
we made a version of the Herman protocol where each participant's bias is
randomly chosen, which ruins the symmetry and so causes the ADD-based approaches
to scale significantly worse (Figures~\ref{fig:expherman13random} and
\ref{fig:expherman17random}, and \ref{fig:expherman19}); we see that symbolic
ADD-based approaches completely fail on Herman 17 and Herman 19 (the curve
terminating denotes a memory error). \rubicon{} and
the explicit approach are unaffected by varying parameters.

\emph{Queues.} The
Queues model has $K$ queues of capacity $Q$ where every step, tasks arrive with a
particular probability. Three queues are of type 1, the others of type 2. We ask the
probability that all queues of type 1 and at least one queue of type 2 is full
within $k$ steps. Contrary to the previous models, the ADD representation of the
transition matrix is  small. Figure~\ref{fig:queues} shows the relative scaling on this model with
$K=8$ and $Q=3$. We
observe that ADDs quickly fail due to inability to concisely represent the
probability vector $\vec{x}$ from \secref{add}. \rubicon{} outperforms explicit model
checking until $h=10$.

\mysubsubsection{Sampling Parametric Markov Chains}
\label{sec:expsampling}
We evaluate performance for the pMC sampling problem
outlined in Sec.~\ref{sec:probstatement}. Table~\ref{tab:sampling} gives for four models the time to construct the BDD and to perform WMC, as well as the time to construct an ADD in \storm{} and to perform model checking with this ADD. Finally, we show the time for \storm{} to compute the solution function of the pMC (with the explicit representation).
The pMC sampling in \storm{} -- symbolic and explicit --
computes the reachability probabilities with concrete probabilities.
\rubicon{}, in contrast, constructs a `parametric' BDD once, amortizing the cost
of repeated efficient evaluation.
The `parametric BDD' may be thought of as a solution function, as discussed in Sec.~\ref{sec:operational}. \storm{} cannot compute these solution functions as efficiently. 
We observe in Tab.~\ref{tab:sampling} that fast parametric sampling is
realized in \rubicon{}: for instance, after a 40s up-front compilation of
the factories example with 15 factories, we have a solution function in factorized form and it costs an order of magnitude less
time to draw a sample.
Hence, sampling and computation of solution functions of pMCs is a major strength of \rubicon{}.

\begin{table}[t]
  \centering
  \caption{Sampling performance comparison and pMC model checking, time in seconds.}
  \vspace{-2mm}
  \scriptsize{
  \begin{tabular}{l|rr|rr|r}
    \toprule
    \textbf{Model}
    & \multicolumn{2}{c|}{\rubicon{}~~\quad} & \multicolumn{2}{c|}{\storm{} (w/ ADD)~~\quad} & \storm{} (explicit) \\
    & build
    & WMC
    & build
    & solve 
    & pMC solving
    \\
    \midrule
    Herman R 13 $(h=10)$ & 3 & $<1$ & 32 & 18 & $>$ 1800 \\
    Herman R 17 $(h=10)$ & 45 & 28 & $>$1800 & - & $>$ 1800\\
    Factories 12 $(h=15)$ & 2 & $<$1 & 59 & 286 & $>$ 1800 \\
    Factories 15 $(h=15)$ & 40 & 4 & $>$1800 & - & $>$ 1800 \\
    \bottomrule
  \end{tabular}
  }
  \label{tab:sampling}
\end{table}

\section{Discussion, Related Work, and Conclusion}
\label{sec:conclusion}




We have demonstrated that the probabilistic inference approach to probabilistic
model checking can improve scalability on an important class of problems.
Another benefit of the approach is for sampling pMCs. These are used
to evaluate e.g., robustness of systems~\cite{DBLP:conf/srds/AflakiVBKS17}, or
to synthesise POMDP controllers~\cite{DBLP:conf/uai/Junges0WQWK018}. Many
state-of-the-art
approaches~\cite{DBLP:conf/tase/ChenHHKQ013,DBLP:conf/cav/DehnertJJCVBKA15,DBLP:conf/tacas/Cubuktepe0JKT20}
require the evaluation of various instantiated MCs, and \rubicon{}
is well-suited to this setting. More generally, support of inference
techniques opens the door to a variety of algorithms for additional queries,
e.g, computing \emph{conditional
probabilities}~\cite{DBLP:conf/tacas/AndresR08,DBLP:conf/tacas/BaierKKM14}.

An important limitation of probabilistic inference is that only finitely many
paths can be stored. For infinite horizon properties in cyclic models, an
infinite set of arbitrarily long paths would be required. However, as standard
in probabilistic model checking, we may soundly approximate infinite horizons.
Additionally, the inference algorithm in \dice{} does not support a notion of non-determinism.
It thus can only be used to evaluate MCs, not Markov decision
processes. However, \cite{DBLP:conf/cav/VazquezChanlatte20} illustrates that
this is not a conceptual limitation.
Finally, we remark that \rubicon{} achieves its performance with a straightforward translation. We are optimistic that this is a first step towards supporting a larger class of models by improving the transpilation process for specific problems.

\mysubsubsection{Related work}
\label{sec:related}
The tight connection with inference has been recently investigated via the use of model checking for Bayesian networks, the prime model in probabilistic inference~\cite{DBLP:journals/corr/abs-2007-15071}. Recently, this has been extended to consider parameter synthesis methods from the verification community~\cite{DBLP:journals/corr/abs-2105-14371}.
Bayesian networks can be described as probabilistic programs~\cite{DBLP:conf/esop/BatzKKM18} and their operational semantics coincides with MCs~\cite{DBLP:journals/pe/GretzKM14}. Our
work complements these insights by studying how symbolic model checking can be sped up
by probabilistic inference. 

The path-based perspective is tightly connected to \emph{factored state spaces}. Factored state spaces are often represented as (bipartite) Dynamic Bayesian networks. ADD-based model checking for DBNs has been investigated in~\cite{DBLP:conf/atva/DeiningerDM16}, with mixed results. Their investigation focuses on using ADDs for factored state space representations. We investigate using BDDs representing paths. 
Other approaches also investigated a path-based view:
The symbolic encoding in~\cite{DBLP:conf/hybrid/FranzleHT08} annotates propositional sub-formulae with probabilities, an idea closer to ours.
The underlying process implicitly constructs an (uncompressed) CT leading to an exponential blow-up. Likewise, an explicit construction of a computation tree without factorization is considered in~\cite{DBLP:conf/vmcai/WimmerBB09}.  Compression by grouping paths has been investigated in two \emph{approximate} approaches:~\cite{DBLP:conf/qest/RabeWKYH14} discretises probabilities and encodes into a satisfiability problem with quantifiers and bit-vectors. 
This idea has been extended~\cite{DBLP:journals/corr/abs-1903-09354} to a PAC algorithm by purely propositional encodings and (approximate) model counting \cite{DBLP:conf/ijcai/ChakrabortyFMV15}. 
Finally, factorisation exploits symmetries, which can be exploited using symmetry reduction~\cite{DBLP:conf/cav/KwiatkowskaNP06}. We highlight that the latter is not applicable to the example in Fig.~\ref{fig:motiv}.

There 
are many techniques for exact probabilistic inference in various forms
of probabilistic modeling, including probabilistic graphical
models~\cite{Pearl1988PRIS,Darwiche2011}. The semantics of graphical models make
it difficult to transpile \prism{} programs, since commonly used operations are lacking. Recently, \emph{probabilistic programming languages} have been
developed which are more amenable to
transpilation~\cite{Carpenter2016,de2007problog,gordon2014probabilistic,vandemeent2018introduction,gehr2016psi}.
We target \dice{} due to the technical development that it enables in
Section~\ref{sec:dd}, which enabled us to design and explain our experiments.
Closest related to \dice{} is ProbLog~\cite{fierens2015inference}, which is also
a PPL that performs inference via WMC; ProbLog has different semantics from
\dice{} that make the translation less straightforward.
The paper~\cite{DBLP:conf/cav/VazquezChanlatte20} uses an encoding similar to \dice{} for inferring specifications based on observed traces.
ADDs and variants have been considered for probabilistic inference
\cite{chavira2007compiling,claret2013bayesian,DBLP:conf/pldi/SmolkaKKFHK019}, which is similar to
the process commonly used for probabilistic model checking. The planning community has developed their own disjoint
sets of methods~\cite{DBLP:journals/jair/KlauckSHH20}. Some ideas from learning
have been applied in a model checking
context~\cite{DBLP:conf/atva/BrazdilCCFKKPU14}.

\mysubsubsection{Conclusion}
We present \rubicon{}, bringing probabilistic AI to the probabilistic model
checking community. Our results show that \rubicon{} can outperform
probabilistic model checkers on some interesting examples, and that this is not
a coincidence but rather the result of a significantly different perspective.


{
\bibliographystyle{splncs}
\bibliography{bibliography}
}

\appendix
\clearpage

\section{Binary MCs}
\label{sec:binary}

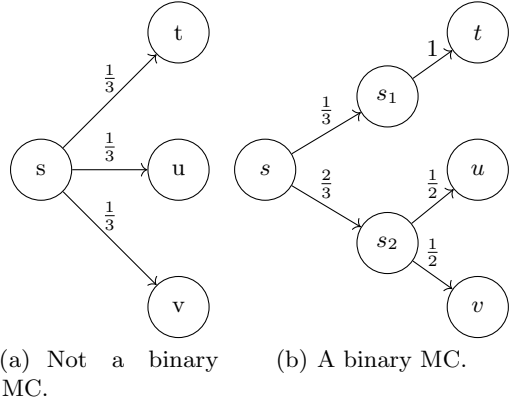
\begin{figure}[t]
\centering
\subfigure[Not a binary MC.]{
\begin{tikzpicture}
	\node[state] (s) {s};
	\node[state,right=of s] (u) {u};
	\node[state,above=of u] (t) {t};
	\node[state,below=of u] (v) {v};
	
    \draw[->] (s) edge node[above] {$\frac{1}{3}$} (t);
    \draw[->] (s) edge node[above] {$\frac{1}{3}$} (u);
    \draw[->] (s) edge node[above] {$\frac{1}{3}$} (v);
\end{tikzpicture}	
}
\subfigure[A binary MC.]{
\begin{tikzpicture}
	\node[state] (s) {$s$};
	\node[state,right=0.8cm of s, yshift=3em] (s1) {$s_1$};
	\node[state,right=0.8cm of s, yshift=-3em] (s2) {$s_2$};
	\node[state,right=2cm of s] (u) {$u$};
	\node[state,above=of u] (t) {$t$};
	\node[state,below=of u] (v) {$v$};
	
	\draw[->] (s) edge node[above] {$\frac{1}{3}$} (s1);
    \draw[->] (s) edge node[above] {$\frac{2}{3}$} (s2);

    \draw[->] (s1) edge node[above] {$1$} (t);
    \draw[->] (s2) edge node[above] {$\frac{1}{2}$} (u);
    \draw[->] (s2) edge node[above] {$\frac{1}{2}$} (v);
\end{tikzpicture}	
}
\caption{Making an MC binary.}
\label{fig:binary}
\end{figure}

Any (parametric) Markov chain with out-degree more than two can translated into a Markov chain with an out-degree of at most two. This operation is standard and e.g. exploited in~\cite{DBLP:conf/uai/Junges0WQWK018}. 
We exemplify one possible construction in Fig.~\ref{fig:binary}. Notice that this construction requires increasing the horizon.

\section{Details on \rubicon: \prism{} to \dice{}}
\label{sec:translation:extensions}
In this section, we assume some familiarity with the \prism{} semantics. We refer to the \prism{} website for details.

\begin{figure}[t]

\subfigure[]{
\usebox{\prismpd}
\label{fig:prismpd}
}
\subfigure[]{
\usebox{\dicepd}
\label{fig:dicepd}
}
\caption{Illustrating state-dependent probabilities}	
\end{figure}

\subsection{Extensions to Monolithic Translation}

\myparagraph{State-dependent probabilities.}
\dice{} currently does not support expressions that evaluate to rationals, and
thus, probabilities are constants. Thus, our translation expands expressions by
considering all values for the variables that occur in these statements. We
illustrate this in Fig.~\ref{fig:prismpd} and Fig.~\ref{fig:dicepd}. \prism{}
programs may contain expressions like $\nicefrac{x}{y}$, with $x$ and $y$ both
ranging from, say, $0$ to $10$ which may not necessarily be
probabilities. The language only requires the outcomes to be valid
probabilities for reachable expressions. 

\myparagraph{Init statements}
The declarative way of initial states -- that give an initial state as the solution of a predicate -- is supported by using \storm's API. Notice that currently, we only support MCs with a unique initial state.

\subsection{Modular Translation}

For modular \prism{} programs, the \emph{action names} at the front of \prism{} commands  are important.  In each module, there is a set of action names available. An action is \emph{enabled} if each module that contains this action name has (at least) one command with this action whose guard is satisfied.
Commands with an empty action are assumed to have a globally unique action name, so in that case the action is enabled iff the guard is enabled.
Intuitively, once an action is selected, we randomly select a command per module in all modules containing this action name.  
We then independently and randomly, according to the probability distribution over updates, select an update, and execute these updates in parallel.
All \readcmd{}s are done before all \writecmd{}s. We remark that variables can
be read from and written to from any module. Data races lead to undefined
behavior, i.e., any linearization of updates is valid in \prism{}.

\myparagraph{Modular translation without overlapping guards within action and module.}
Here, we assume guards do not overlap within an action and module, see the next paragraph for the general case. 
The \dice{} program
first evaluates all actions to determine their joint guards. Then \dice{} randomly selects one of the actions which is enabled\footnote{The semantics can be thought of as applying a uniform scheduler to an underlying MDP where all actions are represented.}. 
Once the action is fixed, we now need to select to associated commands and updates. 
While similar to the vanilla case, we now run a series of updates rather than a single update.
More precisely, once the actions are fixed, we iterate over the modules and flip the coins to select the updates for each command.
We use the outcomes of these coin flips to incrementally construct the next state. 
We remark that the latter is not completely trivial as for different actions, different modules may be assigning an update to a particular variable.\footnote{Recall, \prism{} semantics require that there are no data races.}

\myparagraph{Modular translation with overlapping guards within action and module.}
When a module has multiple commands with the same action, the semantics
of \prism{} programs requires uniform resolving of actions on a
\emph{global} level, among all enabled combinations of enabled
commands.  Thus in our example, if there were two \emph{enabled}
\texttt{a} actions in module \texttt{m1} in a given state,
then \texttt{a} actions would get double weight when determining which action to select.  Once we have computed the
right weight for every action, we
can then continue as before, where we now in every module must first decide
which action to take.

\subsection{Discussion on sampling and other properties}

\myparagraph{Sampling and symbolic distributions.}
As discussed before, \dice{} may quickly evaluate models with a range of distributions. 
Technically, we support \prism{} programs with symbolic probabilities (parameters), and allow probabilities to be expressions over these symbols. 
We collect all commands that depend on these parameters, and replace these by symbolic transitions. We then (separately) translate each assignments to these parameters to concrete instantiations of the symbolic distributions.

\myparagraph{Extended finite-horizon properties.}
Instead of returning the distribution over a predicate whether a target state has been visited, 
the \dice{} program can return distributions over (bounded) quantities. In the finite horizon case, expected cumulative rewards (that assign to every finite path a bounded quantity rather than true or false) can thus be supported straightforwardly. 
Rather than the simple reachability, the target can be straightforwardly described by an automaton. The translation merely needs to update the \texttt{hit} function (and make it a stateful function). 
\dice{} has native and efficient support for conditioning, which allows conditioning over a finite horizon events, e.g., to condition on a prefix, or to condition that within the first $H_1$ steps, a particular state must have been visited. 
Combinations of these constructions with indefinite horizon properties are left for future work.

\myparagraph{Indefinite horizon.}
Inspired by ideas like interval iteration~\cite{DBLP:journals/tcs/HaddadM18} is the following approximation.
Naturally, the probability mass for the bounded horizon is a lower bound on the indefinite horizon probability. The also obtain an upper bound, we use the following equality~\cite{BK08}: \[ \ \Pr(\eventually T) + \Pr(\eventually (\globally \neg T)) = 1, \] that states that eventually we reach a target state, or we reach a state from which it is impossible to reach a target state, denoted $\globally \neg T$ (`globally not $T$'). By setting $(\globally \neg T)$ as the bad states, we can approximate $\Pr(\eventually \text{Bad})$ with a bounded horizon probability, getting 
\[ \Pr(\eventuallyb{h} T) \leq \Pr(\eventually T) \leq  1- \Pr(\eventuallyb{h} \mathrm{Bad}). \]
To generate a \dice{} program, we compute with Storm a  BDD that expresses the states in $\text{Bad}$~\cite{BK08}. We translate this BDD in a sequence of if-then-else statements, with one statement per node.

\lstset{language=Prism}   
\newsavebox{\prismparametric}
\begin{lrbox}{\prismparametric}
\begin{lstlisting}[numbers=none]
module main
x : [0..1] init 0;
y : [0..2] init 1;

const double p,q,u;

[] x=0&y<2 -> p:x'=1 + 1-p:y'=y+1;
[] y=2 -> q*q:y'=y-1 + u:y'=y; 
[] x=1&y!=1 -> 1:x'=y & y'=x;
endmodule
\end{lstlisting}
\end{lrbox}
\lstset{language=dice}   
\newsavebox{\diceparametric}
\begin{lrbox}{\diceparametric}
\begin{lstlisting}[numbers=none]
fun step( s:(x,y) ) {
  if x==0 && y<2 then
   if flipsym p (1,y) else (x,y+1)
  else if y == 2 then 
  if flipsym qu (x,y-1) else (x,y)
  else if x==1 && y != 1 then (y,x)
  else (x,y)
} 
\end{lstlisting}
\end{lrbox}
\lstset{language=dice}   
\newsavebox{\diceadditional}
\begin{lrbox}{\diceadditional}
\begin{lstlisting}[numbers=none]
// p=0.6, q=0.5, u=0.75
p=0.6,qu=0.25
// p=0.3, q=0.1, u=0.99
p=0.3,qu=0.01
// p=0.3, q=0.1, u=0.1
// not valid
\end{lstlisting}
\end{lrbox}

\begin{figure}[t]
    \centering
    \subfigure[]{\usebox{\prismparametric}}
    \subfigure[]{\usebox{\diceparametric}}
    \caption{Symbolic probabilities. $p=0.6, q=0.5, u=0.75$ is mapped to $p=0.6,qu=0.25$, $p=0.3, q=0.1, u=0.99$ to 
$p=0.3,qu=0.01$, and 
$p=0.3, q=0.1, u=0.1$ yields an error.}
\end{figure}

\newsavebox{\selectfrom}
\begin{lrbox}{\selectfrom}
\begin{lstlisting}
fun selectFrom(a,b,c) {
  let N = (a?1:0)+(b?1:0)+(c?1:0) in
  if N == 0 then 0 else 
     let C = uniform(N)           in 
     if a && C == 1 then 1 else
        let C = if a then C - 1 else a - 1 in 
        if b && C == 1 then 2 else   
        3
} 
\end{lstlisting}
\end{lrbox}

\begin{figure}[t]
    \centering
    \subfigure[]{\usebox{\selectfrom}}
    \label{fig:selectfrom}
    \caption{\texttt{selectFrom} auxiliary function}.
\end{figure}

\subsection{Technical details}

\myparagraph{Invalid inputs}
The semantics for 
\prism{} programs assume that bounds are adhered to. However, \rubicon{} does not enforce this.

\myparagraph{Overlapping guards}
To avoid constantly running into the overlapping guards case, we run an Satisfiability-Modulo-Theories~\cite{DBLP:series/faia/BarrettSST09} -solver that checks whether commands have overlapping guards.  
This analysis may be refined, e.g., to take into account for which states we run into overlapping guards.

\myparagraph{selectFrom}
\texttt{selectFrom} is not a native function in \dice{} but rather encoded as in Figure~\ref{fig:selectfrom}. We first count the number of set bits, then select randomly an offset \texttt{C} and then count until we found the $\texttt{C}$'th set bit, and return its index.

\myparagraph{Bitwidth and domains.}
Notice that the translation requires the lower bounds of all variables to be $0$. 
Dice programs type integers in their bitwidth. 
This potentially leads to typing errors when variables with different bitwidths occur within an expression. 
We therefore use the bitwidth of the largest domain for all variables. Static analysis could potentially refine this. 
We do not explicitly check whether variables remain in their domain, the behavior of violating variable bounds is undefined in \prism{} semantics.

\myparagraph{Further technical concerns.}
Furthermore, we have seen problems with expressing that exactly one of a set of predicates $\phi_1,\hdots,\phi_k$ should be true. In Prism programs, this is often expressed with $(\phi_1?1:0)+\hdots+ (\phi_k?1:0)=1$, which is awkward for the aforementioned typing problem. We alleviate this specific problem by extending the Prism dialect that Storm accepts with predicates like $\texttt{ExactlyOneOf}(\phi_1,\hdots,\phi_k)$. 

\section{Additional Experiments}
\label{app:exp}

See Table~\ref{tab:brp}.

\begin{table}[h!]
  \centering
  \begin{tabular}{cccccccc}
    \toprule
    $N$ ~& Max ~& $h$ & \rubicon{} (s) ~& \storm{} Sym. (s) ~& \storm{} Expl. (s) & Transition Size & BDD Size \\
    \midrule
    16 & 3 & 10 & 0.48 & 0.4 & $<$ 0.1 & 1806 & 6\\
    16 & 3 & 40 & 13.89 & 0.4 & $<$ 0.1 & 1806 & 146\\
    16 & 6 & 10 & 0.46 & 0.47 & $<$ 0.1 & 1859 & 2 \\
    16 & 6 & 40 & 16.67 & 1.17 & $<$ 0.1 &  1859 & 209 \\
    32 & 2 & 10	& 0.55 & $<$0.1 & 	0.5	& 1950 &	5			\\
	64 &2	&40	&20.4	&1.07	&$<$0.1	&1950	&113		\\	
	128 &2	&10	&0.73	&0.778&	$<$0.1&	2019&	5			\\
	128&2	&20	&5.25	&1.28	&$<$0.1	&2019	&33		\\	
    \bottomrule
  \end{tabular}
  \caption{Comparisons for \texttt{brp}}
  \label{tab:brp}
\end{table}


\section{Models}
See below.
\pagebreak
\begin{lstlisting}[frame = single, caption = {The ``Weather factory'' factory
\prism{} model with 7 factories.}]
dtmc

const double p1 = 0.1;
const double q1 = 0.2;

const double p2 = 0.2;
const double q2 = 0.3;

const double p3 = 0.41;
const double q3 = 0.45;

const double p4 = 0.94;
const double q4 = 0.243;

const double p5 = 0.434;
const double q5 = 0.293;

const double p6 = 0.4341;
const double q6 = 0.2934;

const double p7 = 0.4345;
const double q7 = 0.2939;


module weathermodule
    sun : bool init true;
    [act]  sun -> 0.7: (sun'=sun) + 0.3: (sun'=!sun);
    [act] !sun -> 0.4: (sun'=sun) + 0.6: (sun'=!sun);
endmodule

module factory1
    state1 : bool init false;
    [act] state1 & sun  -> 0.3 * p1: (state1'=true) + 1-(0.3 * p1): (state1'=false);
    [act] !state1 & sun -> 0.7 * q1: (state1'=true) + 1-(0.7 * q1): (state1'=false);
    [act] state1 & !sun -> 0.6 * p1: (state1'=true) + 1-(0.6 * p1): (state1'=false);
    [act] !state1 & !sun -> 0.4 * q1: (state1'=true) + 1-(0.4 * q1): (state1'=false);
endmodule

module factory2 = factory1[state1=state2,p1=p2,q1=q2] endmodule
module factory3 = factory1[state1=state3,p1=p3,q1=q3] endmodule
module factory4 = factory1[state1=state4,p1=p4,q1=q4] endmodule
module factory5 = factory1[state1=state5,p1=p5,q1=q5] endmodule
module factory6 = factory1[state1=state6,p1=p6,q1=q6] endmodule
module factory7 = factory1[state1=state7,p1=p7,q1=q7] endmodule

label "allStrike" = state1 & state2 & state3 & state4 & state5 & state6 & state7;
\end{lstlisting}

\pagebreak
\begin{lstlisting}[frame = single, caption = {The ``Queues'' \prism{} model.}]
dtmc

const double p1=0.4;
const double p2=0.5;
const double p3=0.65;
const double p4=0.75;
const double p5=0.85;
const double p6=0.9;
const double p7=0.92;
const double p8=0.96;

const int N = 5;
const int N1 = N;
const int N2 = N;
const int N3 = N;
const int N4 = N;
const int N5 = N;
const int N6 = N;
const int N7 = N;
const int N8 = N;

module queue1
    pos1 : [0..N1] init 0;
    [step] pos1 < N1 -> p1: (pos1'=pos1+1) + 1-p1: (pos1'=pos1);
    [step] pos1 = N1 -> 1: (pos1'=pos1);
endmodule

module queue2=queue1[pos1=pos2,p1=p2,N1=N2] endmodule
module queue3=queue1[pos1=pos3,p1=p3,N1=N3] endmodule
module queue4=queue1[pos1=pos4,p1=p4,N1=N4] endmodule
module queue5=queue1[pos1=pos5,p1=p5,N1=N5] endmodule
module queue6=queue1[pos1=pos6,p1=p6,N1=N6] endmodule
module queue7=queue1[pos1=pos7,p1=p7,N1=N7] endmodule
module queue8=queue1[pos1=pos8,p1=p8,N1=N8] endmodule


label "target" = pos1=N1 & pos2=N2 & pos3=N3 & (pos4 < N4 | pos5 < N5 | pos6 < N6 | pos7 < N7 | pos8 < N8);
\end{lstlisting}


\end{document}